\documentclass[aps,prb,twocolumn]{revtex4-1}
\usepackage{graphicx}
\usepackage{float}
\usepackage{times}
\usepackage{color}
\input pdfcolor.tex
\usepackage{graphicx}
\usepackage{epsfig,color}

\begin{document}

\author{Sauri Bhattacharyya$^{1}$, 
	Saurabh Pradhan$^{1,2}$ and Pinaki Majumdar$^{1}$}

\affiliation{
	$^1$~Harish-Chandra Research Institute, HBNI,
	Chhatnag Road, Jhunsi, Allahabad 211 019, India\\
	$^2$~Department of Physics and Astronomy, Uppsala University,
	751 05 Uppsala, Sweden
}

\title{Modeling dynamical phonon fluctuations across the magnetically 
driven polaron crossover\\
 in the manganites}

\date{\today}

\begin{abstract}
We investigate the dynamical structure factor associated with lattice
fluctuations in a model that approximates the manganites. It involves
electrons strongly coupled to core spins, and to lattice distortions,
in a weakly disordered background. This model is solved in the adiabatic
limit in two dimensions via Monte Carlo, retaining all the thermal 
fluctuations. In the metallic phase near the polaronic crossover this 
approach captures the effect of thermally induced polaron formation, 
and their short range correlation, on the electronic spectral functions.
The dynamical fluctuations of the optical and acoustic phonon modes are 
computed at a `one loop' level by calculating the electronic polarisability 
in the the thermally fluctuating backgrounds, and solving the phonon 
Dyson equations in real space. We present phonon lineshapes across the 
ferromagnet to paramagnet thermal transition and correlate them with 
changing electronic properties. We compare our results with inelastic 
neutron scattering data on the metallic manganites, and also predict 
what one may find in the more insulating phases. 
\end{abstract}


\maketitle

\section{Introduction}

The dispersion of phonons in metals has a `direct'
part, determined by the interactions between the ions,
and an indirect part determined by electron-phonon 
coupling and the electronic polarisability \cite{phon}.
The damping similarly can arise from phonon-phonon
interaction - related to nonlinearities in the
inter-ionic potential, or the polarisability
- via particle-hole pair creation. Phonon physics
associated with ionic interactions is well 
understood but the effects arising out of 
a non trivial polarisability, in correlated
electron systems, have generated a new list of
questions.

Many electronic systems have now been shown
to exhibit behaviour that differs widely 
from the band picture 
\cite{dag}. In particular correlation
effects give rise to competing ordered phases,
short range order, and sometimes a pseudogapped
electronic spectrum - widely at variance with 
band theory\cite{man,cup}. The short range order
gives a characteristic momentum dependence to
electronic properties, the pseudogap generates
a new frequency scale, and temperature plays 
an important role in the evolution of both these
features. 
The electronic polarisability $\Pi({\bf q},\Omega, T)$,
where ${\bf q}$ is the momentum, $\Omega$ the frequency,
and $T$ the temperature, becomes a non trivial function
- very distinct from its weak coupling Lindhard counterpart.
It is no wonder that phonons in these correlated systems
show a very rich behaviour.

For example, recent experiments 
\cite{weber1,weber2,weber3} reveal 
the impact of short-range              
charge order (CO),  and an electronic pseudogap,
on the softening and broadening of the
bond stretching phonons in the manganites 
for momenta near the
CE ordering wavevector.
Similar effects on phonons have
been observed in underdoped 
cuprates\cite{reznik,dean},
possibly owing to stripe order.
In fact phonon properties near
${\bf q} \sim (\frac{1}{4},\frac{1}{4},0) $,
the CE ordering wavevector, in manganites
are very similar 
to that in cuprates at
${\bf q} \sim (\frac{1}{4},0,0)$. 
In the 
nickelates\cite{kajimoto}, inelastic
neutron scattering reveals strong softening and
branch-splitting of longitudinal optical (LO) phonons.
Strong phonon renormalizations
have also been reported in doped bismuthates\cite{braden}
possibly due to their coupling to interplanar
charge fluctuations.

There are two distinct scenarios to consider when
approaching these problems.
(a)~Situations
where the electron correlations 
arise from inter-electron interactions - and the phonons
are only weakly coupled to the electrons, and 
(b)~where electron-phonon interaction is itself
responsible (if only in part) for features in the
electronic spectrum. Category (a), possibly appropriate
to the cuprates\cite{cup}, requires a solution of the
interacting electron problem and then use of the
resulting $\Pi({\bf q}, \Omega)$ to determine
the phonon self-energy. Category (b) requires an
``ab initio'' handling of lattice distortions,
and is pertinent to the manganites\cite{man}. 

This paper
is focused on understanding phonons in a model
that approximates the manganites. We will consider
models in category (a) elsewhere.
To set the stage we quickly recapitulate the
electronic observations in the intermediate
coupling manganites, and the recent results 
on phonon spectra.

The manganites A$_{1-x}$B$_x$MnO$_3$,
where A is trivalent and B is divalent, 
involve $e_g$ electrons Hund's coupled to $t_{2g}$
based core spins, and to Jahn-Teller (JT) distortions
of the MnO$_6$ octahedra\cite{man,millis1}.
The Hund's coupling promotes ferromagnetism
(and extended electronic states) while the JT 
coupling favours polaron formation (and electron
localisation). The bandwidth and carrier density 
can be tuned via choice of the A and B ions, and
$x$.
A material that is a ferromagnetic metal (FM) 
 at low
temperature can show polaronic signatures with
increasing temperature as spin randomness reduces
the effective bandwidth.  
In fact it can cross over to a paramagnetic
insulator (PI) across the magnetic transition.
There is naturally a drastic change in the 
electronic spectra with temperature in this
parameter window. 

% -----------------------------------------------------------
\begin{figure}[b]
\centerline{
\includegraphics[width=6cm,height=5.4cm]{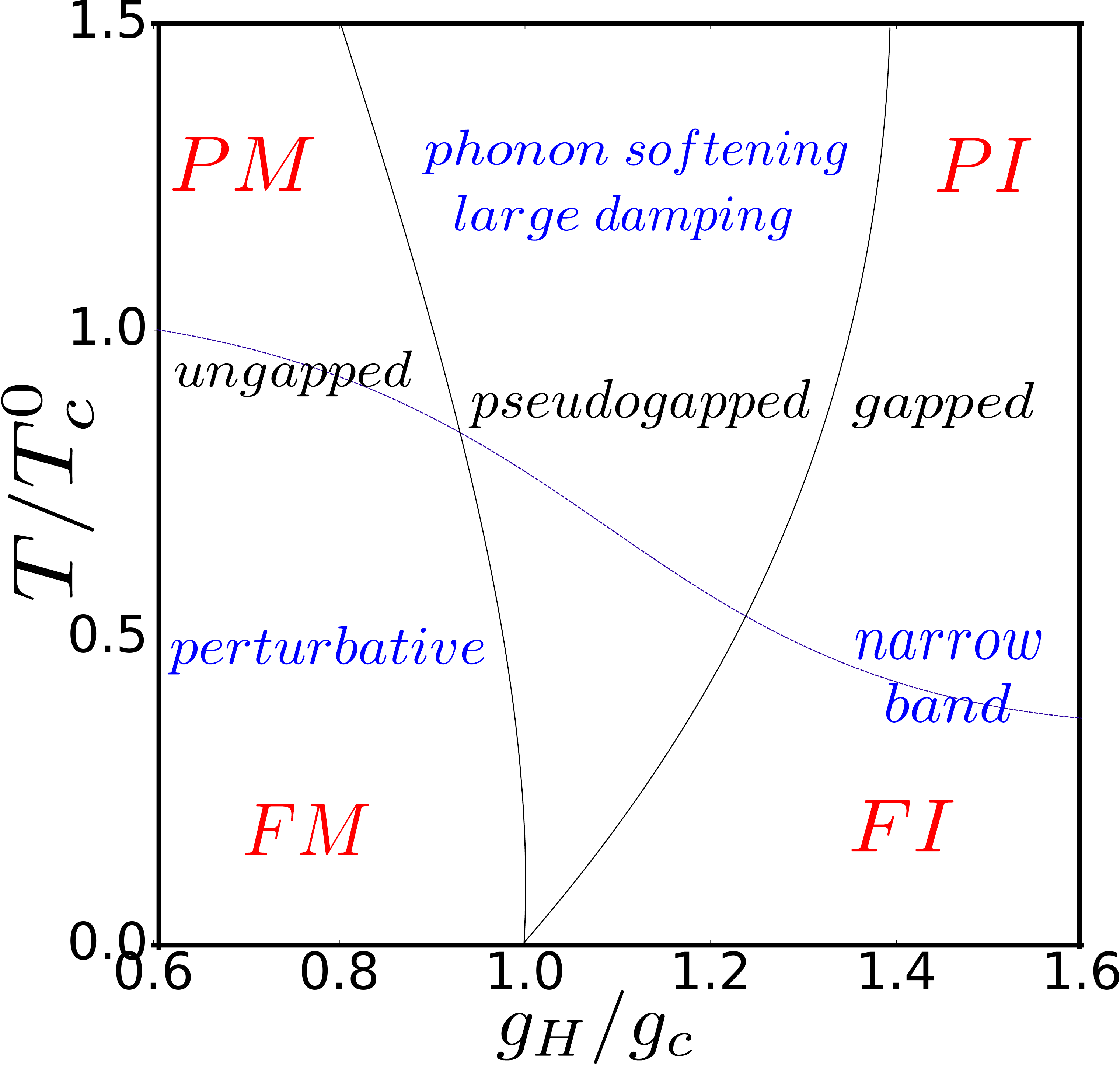}
}
\caption{Phonon `phase diagram' for the 2D
        double exchange Holstein model near half filling
        for varying coupling and temperature.
        We normalize coupling $g_{H}$ by the zero temperature
        critical coupling for polaron formation
        $g_c$ and temperature by the ferromagnetic 
        transition temperature at weak coupling 
        $T^{0}_{c}=2\Omega_{0}$, where $\Omega_{0}$
        is the bare Holstein frequency.
        FM and PM denote
        the ferromagnetic and paramagnetic metal phases,
        similarly FI and PI are ferro and para insulating
        phases. The phonon regimes are indicated in blue.}
\end{figure}
%------------------------------------------------------------

Inelastic neutron scattering
has uncovered the following phonon features
in this regime.
\begin{enumerate}
\item The 
Mn-O bond stretching phonons along the (110)
direction in the bilayer manganite
La$_{1.2}$Sr$_{1.8}$Mn$_{2}$O$_{7}$ soften and broaden
close to the CE ordering wavevector in the low
temperature ($\sim10$K) metallic phase\cite{weber1}.
\item 
A similar effect is observed in the 3D compound
La$_{0.7}$Sr$_{0.3}$MnO$_{3}$ for the transverse
acoustic phonons, on heating through the Curie 
temperature $T_c\sim350$K \cite{weber2}. 
\item 
In the charge and orbital ordered bilayer
manganite LaSr$_{2}$Mn$_{2}$O$_{7}$, one sees
a reduction in broadening and softening of low-energy
acoustic phonons on cooling below $T_{CO} \sim225$K
\cite{weber3}.  
\end{enumerate}

We wish to address these issues, by first solving the 
appropriate electronic problem.
To retain the key qualitative features of manganite
physics, and also to access large spatial scales, we
study a two dimensional model of electrons Hund's 
coupled to core spins and to lattice distortions via
a Holstein coupling. 
We retain weak disorder in the model - to mimic the
effect of cation disorder in the doped materials. 
This problem is solved via a Monte Carlo method in
real space, and the electronic polarisability that
emerges is used to compute the phonon spectra.

We list our main results below. In our
notation $g_H$ is the Holstein
coupling and $g_c$ its value 
for the adiabatic polaron transition.
\begin{enumerate}
\item 
At $g_{H} \ll g_c$ 
the ground state is a ferromagnetic metal 
at low temperature, with dispersive phonons that show
a Kohn anomaly close to half filling. With increasing
temperature spin disorder and thermally induced lattice
distortions create a a mild electronic pseudogap, and
also  broaden the  phonons considerably.
\item 
For $g_{H} \lesssim g_c$ the interplay of electron-phonon coupling
and disorder localises the phonons.  
The phonon spectrum, as a result, is quite
incoherent. The electronic density of states features
a pseudogap. Thermal effects further accentuate
lifetimes.
\item 
For $g_{H} \gtrsim g_c$ the ground state is a 
ferromagnetic insulator. 
The electronic spectrum is gapped 
and phonon band is narrow, with
bandwidth limited lifetimes. This is related to
emergent short-range order in the background. On
heating past the ferromagnetic $T_c$, phonon lifetimes 
pick up.
\end{enumerate}

Fig.1 summarizes our findings regarding thermodynamic
phases (indicated in red), electronic states (in black) 
and phonon regimes (in blue).
It's discussed in detail later in context of the
thermal physics. In reporting the results, we first
focus on the ground state. The next section exhibits
our findings about the thermal physics. This is followed
by the discussion section, summarizing the results and
making a qualitative correspondence with experiments,
and a concluding section.

\section{Model and method}

We study the disordered Holstein-double exchange (d-HDE)
model, coupled to acoustic phonons on a two-dimensional 
square lattice. The full problem has three parts- 
(i)~the `d-HDE' part - electrons
coupled to optical phonons and `core' spins,
in a weakly disordered background, 
(ii)~an acoustic phonon mode, and (iii)~`weak' 
coupling between the acoustic 
phonons and electrons. The total Hamiltonian is:
\begin{equation}
H_{tot}=H_{1}+H_{2}+H_{3}
\end{equation}
The Hamiltonian for the d-HDE part is- 
\begin{eqnarray}
	H_{1} &=& \sum_{<ij>\sigma} t_{ij}c^{\dagger}_{i\sigma}c_{j\sigma} 
	-J_{H}\sum_{i}\vec{S}_{i}.\vec{\sigma}_{i}  
	+ \sum_{i}(\epsilon_{i}-\mu)n_{i} \cr
        && ~~-g_{H}\sum_{i}n_{i}x_{i} 
         +\sum_{i}(\frac{p^2_{i}}{2M} + \frac{1}{2}Kx^2_{i})
\end{eqnarray}

Here, $t_{ij}$'s are the hopping amplitudes. We study a 
nearest neighbour model 
with $t=1$ for a nearly commensurate density, viz. 
$n=0.40$.
$K$ and $M$ are the stiffness constant and mass, respectively,
of the optical phonons, 
and $g$ is the electron-phonon coupling constant. 
We set $K=1$. In this paper, we 
report studies for $\Omega_0 = \sqrt{K/M} = 0.05$, 
which is a reasonable value for real materials. 
$\vec{S}_{i}$'s are `core spins', assumed
 to be large and classical.
$\epsilon_{i}$ is a quenched binary disorder with zero mean
and value $\pm \Delta$. 
The chemical potential $\mu$ is varied to maintain 
the electron density at the required value. We'll work in
the extreme Hund's coupling limit $J_{H}/t\rightarrow \infty$.
This effectively enslaves the electrons spins to the core spin
orientation and one obtains an effectively spinless, hopping
disordered model- 
\begin{eqnarray}
H_{1}&=&\sum_{<ij>}t_{ij}{\gamma}^{\dagger}_{i}{\gamma}_{j} 
+\sum_{i}(\frac{p^2_{i}}{2M} + \frac{1}{2}Kx^2_{i}) \nonumber \\
&+& \sum_{i}(\epsilon_{i}-\mu-g_{H}x_{i})n_{i}
\end{eqnarray}
The hopping amplitude depends on spin 
orientation via the relation-
\begin{equation}
t_{ij}/t=cos\frac{\theta_i}{2}cos\frac{\theta_j}{2}
 + isin\frac{\theta_i}{2}sin\frac{\theta_j}{2}e^{i(\phi_i-\phi_j)}
\end{equation}
where the $\theta_i$ and $\phi_i$ are, respectively,
 the polar and azimuthal angles of the core spins $\vec{S}_{i}$.
The Hamiltonian for the bare acoustic phonons is-
\begin{equation}
H_{2}=\sum_{\vec{q}}\omega_{\vec{q}}a^{\dagger}_{\vec{q}}a_{\vec{q}}
\end{equation} 
with the bare dispersion 
given by the equation-
\begin{equation}
\omega^2_{\vec{q}}=\frac{2\kappa}{\mu}(2-cos(q_x)-cos(q_y))
\end{equation}
We choose $\kappa=1.0$ and $\mu=2500$ as 
parameters. This ensures that the maximum 
of the acoustic branch is comparable 
to the bare optical 
mode frequency at the zone boundary. 

The coupling between the acoustic phonons 
and electrons is the third part of the Hamiltonian-
\begin{equation}
H_{3}=ig_{ac}\sum_{\vec{k}\vec{q}}\frac{|\vec{q}|}{\sqrt{2\mu\omega_{q}}}
(a_{\vec{q}}-a^\dagger_{-\vec{q}})c^{\dagger}_{\vec{k}+\vec{q}}c_{\vec{k}}
\end{equation}
We've written the parts $H_{2}$ and $H_{3}$ in momentum 
space for simplicity, but the actual calculations using 
them have been implemented in real space. This has the 
advantage of capturing spatial correlations amongst 
static distortions and the 
core spin angles at finite temperature.

We solve the problem hierarchically. First, the d-HDE 
problem is tackled in the adiabatic regime (assuming 
$\Omega_0/t  \rightarrow 0$). 
The $H_{1}$ part in the model leads to the action
\begin{eqnarray}
S &= &\int_0^{\beta}d\tau[
\sum_{ij} \bar{\psi_i}
\{(\partial_{\tau}+\epsilon_i-\mu)\delta_{ij} - t_{ij} \} \psi_j
\cr
&&~~+ \sum_i\bar{\Phi}_i(\partial_{\tau}+\Omega_0) \Phi_i   
-~ g_{H}\sqrt{\frac{\Omega_0}{2K}} 
\sum_{i}(\bar {\Phi}_{i}+\Phi_{i}){\bar\psi_{i}}\psi_{i}]
\nonumber
\end{eqnarray}
Here $\psi_{i}$ and $\phi_{i}$ are the fermion 
and coherent state Bose 
fields respectively.  
$\beta$ denotes the inverse 
temperature, and we use units where
$k_{B}=1$, $\hbar=1$. The relation
between the real and coherent state fields is
\begin{equation}
x_{i}=\sqrt{\frac{1}{2M\Omega_0}}(\Phi_{i}+\bar{\Phi}_{i})
\end{equation}

In quantum Monte Carlo (QMC) 
\cite{QMC}, one `integrates 
out' the fermion fields
to construct the effective bosonic action. The
equilibrium configurations $\Phi_i(\tau)$ 
of that theory are obtained via 
sampling.  Physical correlators are then 
computed as averages with
respect to these $\Phi_i(\tau)$ 
configurations. Within DMFT  
\cite{DMFT} one maps the 
original action in Eq.(2) to an 
impurity problem, with parameters determined
through a self-consistency condition.

In the 
adiabatic limit, or using the static path 
approximation (SPA) strategy,
one treats the $x_{i}$ as classical 
fields, neglecting their imaginary 
time dependence. This corresponds to 
$M \rightarrow \infty$ for a fixed $K$. In addition, the $\theta_i$ 
and $\phi_i$ angles also fluctuate at finite 
temperature. Our method retains the quantum character 
of lattice vibrations perturbatively. 

The full action containing the electrons,  
displacement fields and core spins can be 
rewritten in frequency space as-
\begin{eqnarray}
S &=
&S_{ph} + S_{f}\nonumber \cr
S_{ph}&=&\frac{1}{2}\sum_{i,m}\bar x_{im}(M\omega_{m}^{2}+K)x_{im} \cr
S_{f} &=&\sum_{i,j,\alpha,\beta}\bar{\psi}_{i\alpha}
[(-i\omega_{\alpha}+\epsilon_i-\mu)\delta_{ij}\delta_{\alpha\beta}
- t_{ij}\delta_{\alpha\beta}  \cr
&&~~~~~~~~~~~~~~~~~~-g_{H}\sqrt{\beta}x_{i,\alpha-
	\beta}\delta_{ij}]\psi_{j\beta}
\end{eqnarray} 
where $\alpha$ and $\beta$ are fermionic
 Matsubara frequencies, $m$ is 
a Bose frequency. To get the partition function, 
one integrates over $\psi_{i\alpha}$, $x_{i0}$ and the 
$x_{im}$ fields and of course, 
the angular variables hidden in $t_{ij}$.

The next step is to separate the Bose 
field into zero 
and non-zero Matsubara modes. 
We write $S=S_0 + S_1$ where 
the first part contains fermions
 coupling only to the 
zero frequency (static) mode and
 the second, the finite modes.
We can formally `diagonalize' the
 fermions in presence of the 
static mode to write $S_0$ as-
\begin{equation}
S_{0}={\sum_{l,\alpha}\bar{\xi}_{l,\alpha}
	(-i\omega_{\alpha}+\epsilon_l)\xi_{l,\alpha}} + 
\frac{1}{2}K\sum_ix^2_{i0}
\end{equation}
where $\xi$'s correspond to 
the fermionic eigenmodes in the 
$\{x_{i0} ,\theta_i,\phi_i\}$ background. 
$S_0$ defines the static path 
approximation (SPA) action.

The eigenvalues ($\epsilon_l$'s ) 
depend  non-trivially on the 
$\{x_{i0} ,\theta_i,\phi_i\}$ background. 
To obtain the effective zero mode 
distribution $P\{x_{i0},\theta_i,\phi_i\}$, 
one has to integrate out the 
fermions. At the SPA level,
 there exists an effective 
Hamiltonian $H_{eff}$, 
depending on $\{x_{i0},\theta_i,\phi_i\}$, 
for the fermions and the 
distribution can be formally written as-
\begin{eqnarray}
H_{eff} & =& \sum_{<ij>}t_{ij}
{\gamma}^{\dagger}_{i}{\gamma}_{j} -
 g \sum_{i}n_i x_{i0} \cr
P\{x_{i0},\theta_i,\phi_i\} &= & 
Tr_{{\gamma},{\gamma}^{\dagger}}e^{-\beta (H_{eff} +
	{1 \over 2} K x_{i0}^2) }
\nonumber
\end{eqnarray}

Around the SPA action, we set up a
 cumulant expansion of $S_1$.
Owing to the disparate timescales of the
 optical phonons and electrons, 
we attempt a sequential integrating out. First, 
the fermions are traced out and one obtains 
the following expression for $S_1$
in terms of the $x_i(i\Omega_m)$:
\begin{eqnarray}
S_1 &=
&S^{\prime}_{ph}-Trln[\beta
((-i\omega_{\alpha}+\epsilon_i-\mu)\delta_{ij}\delta_{\alpha\beta} \nonumber 
\\  &&~~~~~~~~~~~~~~~ -t_{ij}\delta_{\alpha\beta}-g_{H}\sqrt{\beta}x_{i,\alpha-
	\beta}\delta_{ij})]\nonumber\\
S^{\prime}_{ph}&=&\frac{1}{2}\sum_{i,m\neq 0}\bar
x_{im}(M\omega_{m}^{2}+K)x_{im}
\end{eqnarray} 

% -----------------------------------------------------------
\begin{figure*}[t]
\centerline{
\includegraphics[height=6cm,width=16cm]{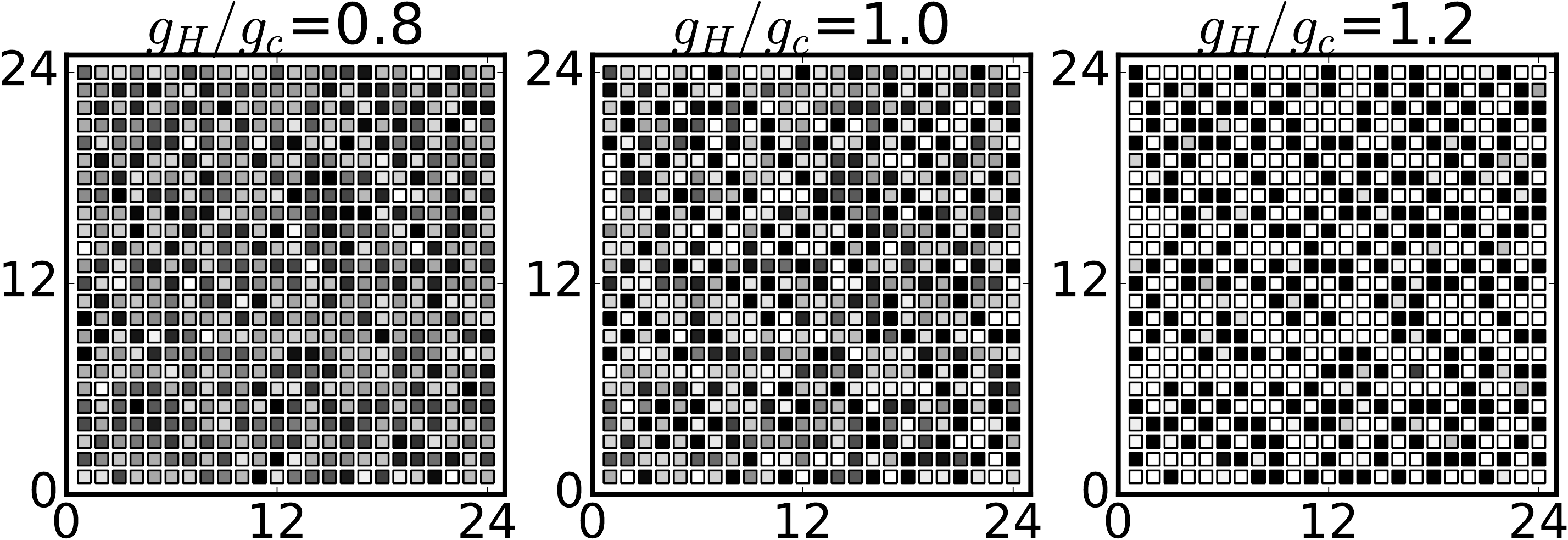}
}
\vspace{.5cm}
\centerline{
\includegraphics[height=4.5cm,width=5.3cm]{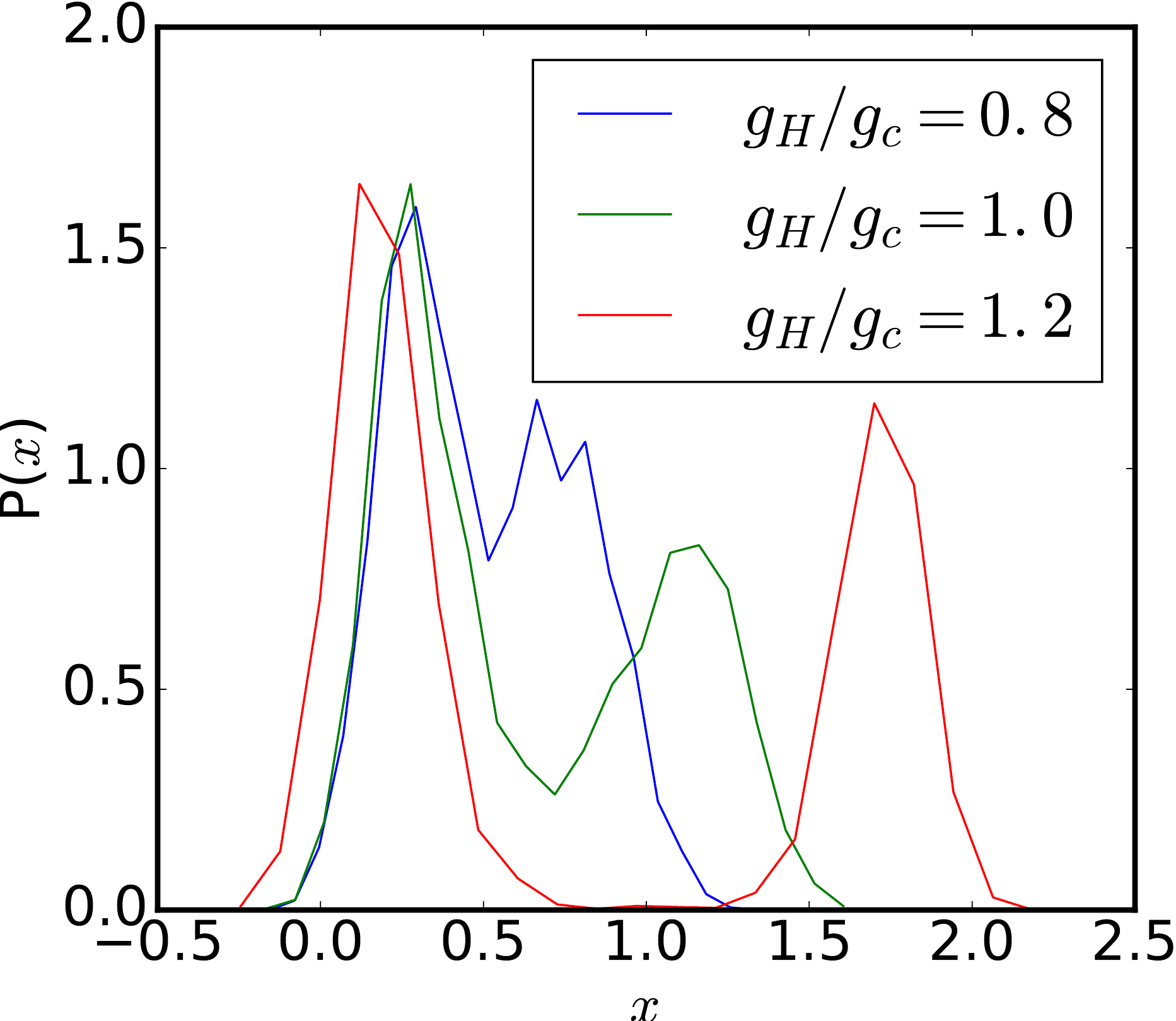}
\includegraphics[height=4.5cm,width=5.3cm]{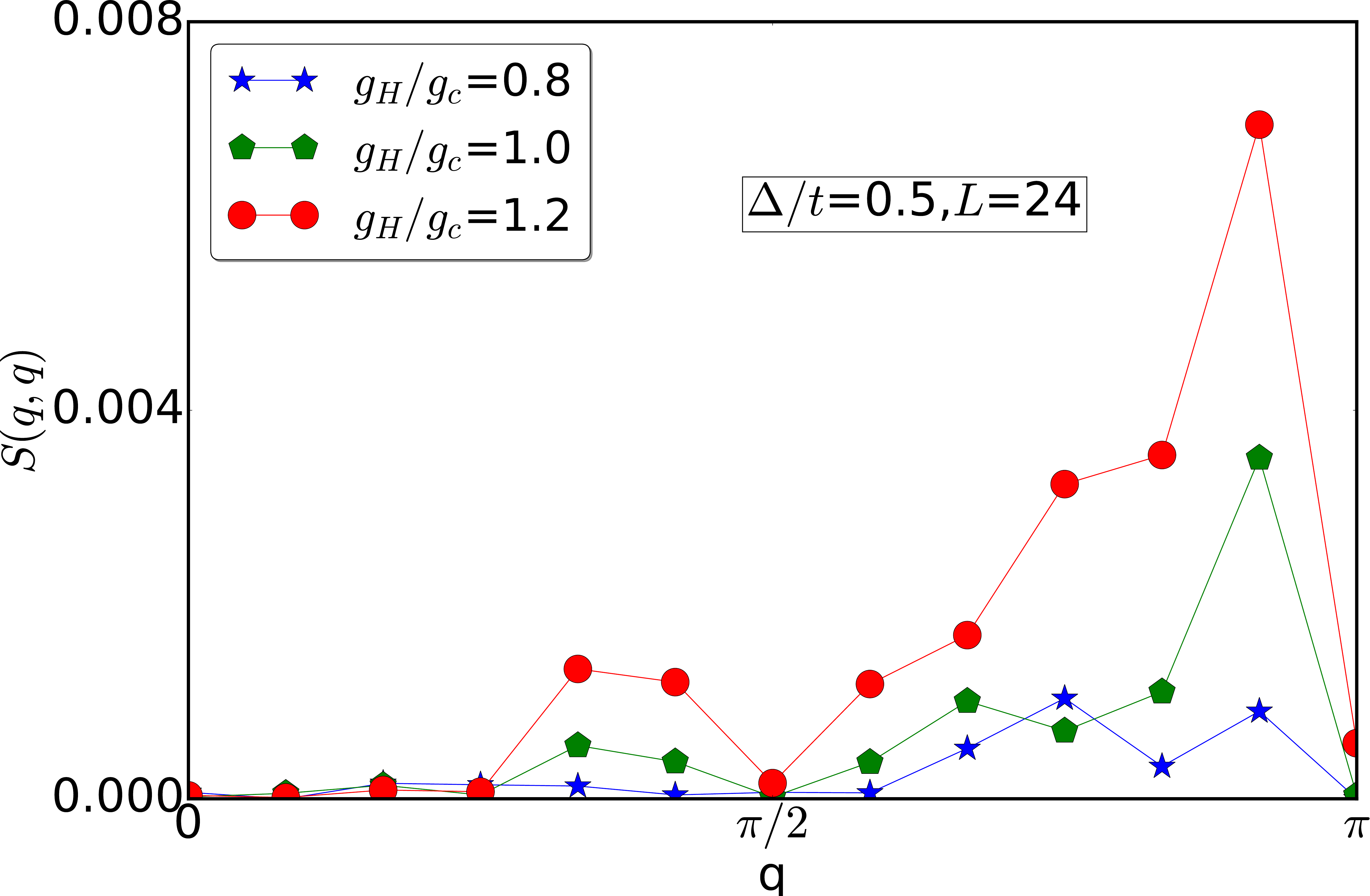}
\includegraphics[height=4.5cm,width=5.3cm]{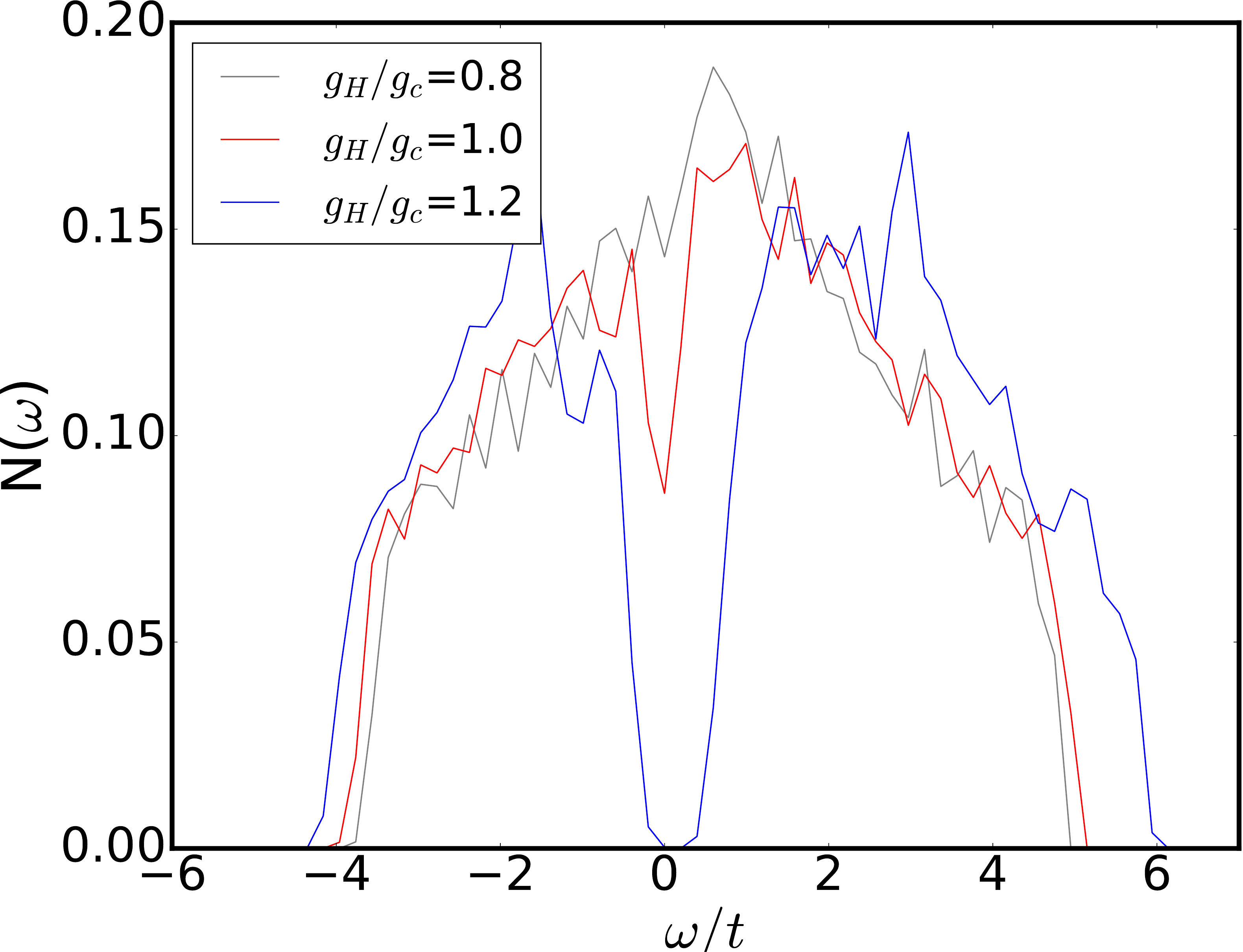}
}
\caption{Top row: Snapshots of static optical phonon
        backgrounds in the ground state for weak ($g/g_c=0.8$,
        left panel), intermediate ($g/g_c=1.0$, middle panel)
         strong ($g/g_c=1.2$, right panel) couplings. We see a
          gradually developing short-range ($\pi,\pi$)
          correlation on moving from left to right, bottom
          row: a) left panel- Distribution of static distortions
          $P(x)$ for three couplings.
We see a weak bimodality transforming to a prominent two-peak distribution at
          strong coupling, b) middle panel- static phonon
          structure factor along the BZ diagonal $S(q,q)$
          for the same coupings, confirming features seen
          in lattice backgrounds, c) right panel-
          electronic density of states $N(\omega)$ from
          weak to strong coupling. We see an evolution from
          tight-binding like spectrum to a small hard-gap DoS
          at strong coupling through a pseudogap regime.
}
\end{figure*}
% -----------------------------------------------------------

Analytic determination of the trace is impossible beyond
weak coupling. 
Our method constitutes of 
expanding the trace in finite
 frequency $x_{i}$ modes upto Gaussian 
level. Physically, we assume
 that the quantum fluctuations are small 
in amplitude, controlled by a
 low $\gamma=\Omega_0/t$ ratio. 
Retaining only 
quadratic terms in the dynamic
 modes allows us to analytically 
integrate them out later. 
An analogous strategy
 has been explored within DMFT before\cite{millis2}. 

 The method is obviously perturbative in $\gamma$ 
 as we diagonalize the fermions 
 in presence of static 
distortions and evaluate 
correlation functions on this `frozen' 
background, which enter
 as coefficients of the 
 quadratic Bose term 
for finite frequencies. After 
re-exponentiating this term 
(in the Gaussian 
approximation spirit), the new partition
function becomes- 
\begin{equation}
Z = \int [D\bar{x}][Dx][D\theta]
[D\phi][D\bar{\xi}][D\xi]e^{-(S_{0}+S_{1})}
\end{equation}
where 
\begin{eqnarray}
S_{0}&=&{\sum_{l,\alpha}\bar{\xi}_{l,\alpha}
	(-i\omega_{\alpha}+\epsilon_l)\xi_{l,\alpha}} 
	+\frac{1}{2}Kx_{i0}^2\\ \nonumber
S_{1}&=&{\sum_{i,j,m>0}\bar{x}_{im}[(M\omega^2_{m}
	+K)\delta_{ij}+{g_{H}^2}
	\Pi_{ij}^{m}(\{x_{i,0},\theta_i,\phi_i\})]x_{jm}}\\ \nonumber
\end{eqnarray}

$\Pi_{ij}^{m}$ is a one-loop
 fermion polarization correcting the 
free Bose propagators. The expression for this is
\begin{equation}
\Pi_{ij}^{m}(\{x_{i0},\theta_i,\phi_i\})=
\frac{1}{\beta}\sum_{\alpha}G_{ij}^{m}G_{ji}^{\alpha-m}
\end{equation} 
$G_{ij}^{m}$'s are Matsubara components
 of real-space fermion Green's 
functions computed in an arbitrary static
 background. One can write a spectral 
representation of this as follows-
\begin{equation}
G^m_{ij}(\{x_{i0},\theta_i,\phi_i\})= 
\sum_n \frac{u_{in}\bar{u}_{jn}}{i\omega_m-\epsilon_n}
\end{equation}
where $u_{in}$ is amplitude at site $i$ for the 
$n$th eigenstate of 
the SPA Hamiltonian and $\epsilon_n$'s are 
the corresponding eigenvalues. 

% -----------------------------------------------------------
\begin{figure*}[t]
        \centerline{
                \includegraphics[width=16cm,height=9cm]{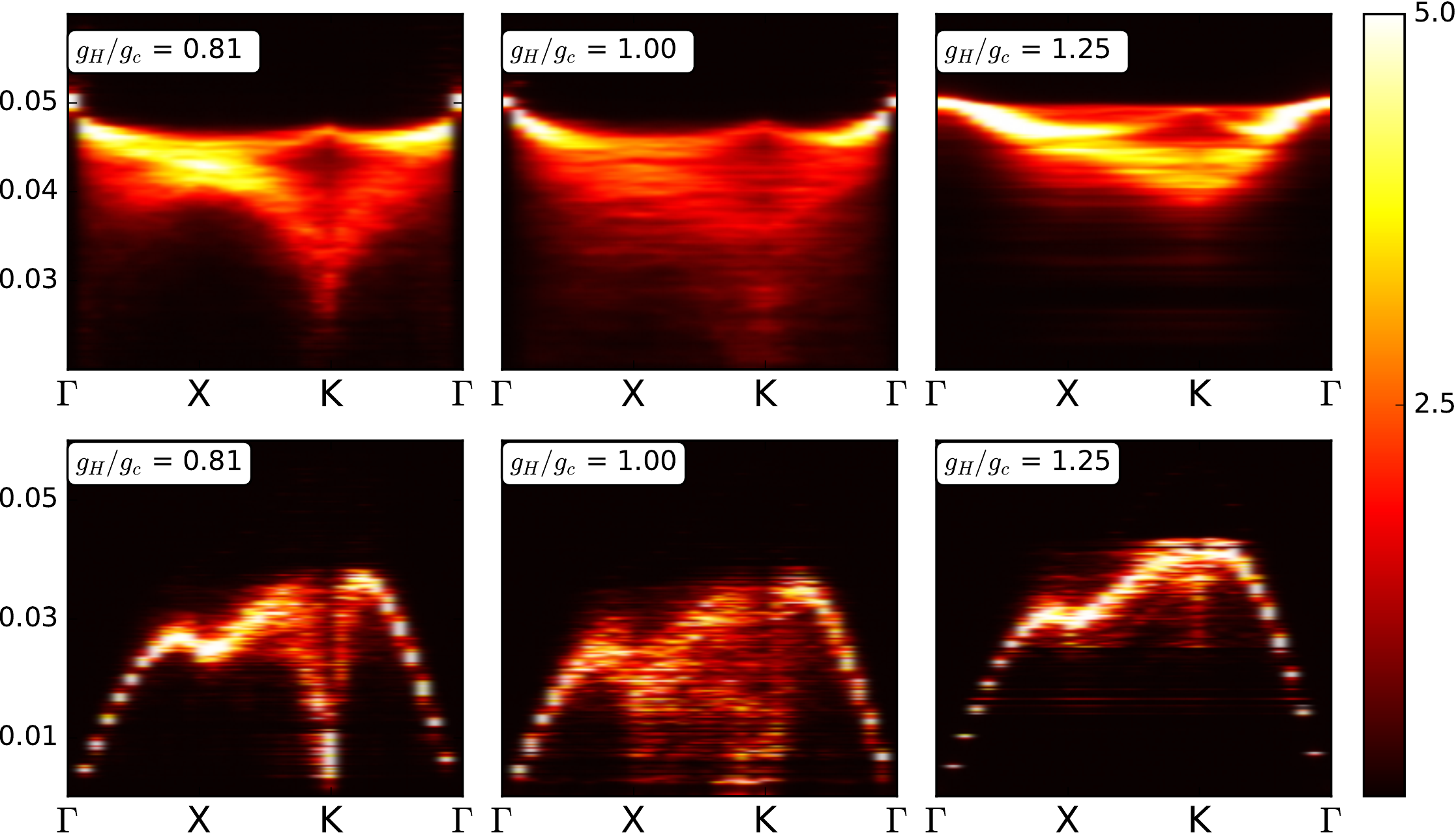}
        }
        \caption{ Spectral distribution of the optical (top panel)
                and acoustic (bottom panel)
                phonon Green's function for varying
                $g_{H}$ in the ground state. The trajectory
                chosen along BZ is $(0,0)\rightarrow(\pi,0)\rightarrow(\pi,\pi)\rightarrow
                (0,0)$. Coupling increases along the horizontal axis in the following sequence
                (from left to right)- $g_{H}/g_{c}=(0.81, 1.00, 1.25)$.
                The critical coupling at this density is $g_{c}=1.60$.
                For the acoustic plots, $g_{ac}=g_{H}$.}
\end{figure*}
%------------------------------------------------------------

If we integrate out the finite 
frequency Bose modes (in $S_1$), 
a new $x_{i0}$ theory emerges.
One can treat this 
resulting action with classical Monte Carlo (MC) 
simulations to obtain the 
optimal background in which the fermions move. 
This basically means 
that at each MC update, one uses the sum of the
SPA free energy and 
the contribution from the finite frequency 
Gaussian fluctuation,
to compute the update probability.

The coefficient of the quadratic Bose 
term of $S_1$ in (13) defines the inverse 
propagator $(D^{-1})_{ij,m}$ for the
 renormalized optical phonons. One can
  view the $\Pi^{m}_{ij}$ as a self-energy 
  for the phonons in real space, 
  correcting the bare propagator. To 
get the renormalized Green's function, 
one needs to solve a Dyson's equation 
on the lattice with possibly inhomogeneous backgrounds.
\begin{equation}
[D]^{-1}_{ij}(\omega)=
[D]^{-1}_{0,ij}(\omega)+g_{H}^{2}[\Pi]_{ij}(\omega)
\end{equation}
Here the bare phonon
 propagator $D^{-1}_{0,ij}$ is defined 
through the equation- 
\begin{equation}
[D]^{-1}_{0,ij}(\omega)=(M\omega^2-K)\delta_{ij}
\end{equation}
where we have analytically 
continued the Green's functions to real
frequency and $D^{-1}_{0,ij}$ 
denotes the inverse propagator for
the bare phonon.
This involves 
a calculation that grows as $O(N^4)$ 
with increasing lattice size $N$. For 
lattices of size 24$\times$24, this 
computation was implemented.

Next, we consider the acoustic phonons. 
As commented earlier, we rewrite the parts 
$H_{2}$ and $H_{3}$ in real space-
\begin{eqnarray}
H_{2}&=&\sum_{ij}M_{ij}a^{\dagger}_ia_j \nonumber \\
H_{3}&=&\sum_{ij\vec{q}}
\frac{ig_{ac}|\vec{q}|}{\sqrt{2\mu\omega_{\vec{q}}}}
(a_je^{i(\vec{r}_i-\vec{r}_j).\vec{q}}-h.c.)c^{\dagger}_ic_i
\end{eqnarray}

The bare phonon is found 
by just inverting the 
matrix $M_{ij}$ in $H_{2}$.
 Its explicit form is given by- 
\begin{equation}
[\tilde{D}]_{0,ij}(\omega)=\frac{e^{-i(\vec{r}_i-\vec{r}_j).
\vec{q}}}{\omega-\omega_{\vec{q}}+i\eta}-
\frac{e^{-i(\vec{r}_i-\vec{r}_j).\vec{q}}}
{\omega+\omega_{\vec{q}}-i\eta}
\end{equation}
where $\eta$ is a positive infinitesimal. 
The coupling $H_{3}$ gives a self-energy 
contribution, which can be 
written as a matrix product-

% -----------------------------------------------------------
\begin{figure}[b]
\centerline{
\includegraphics[height=8cm,width=8.5cm]{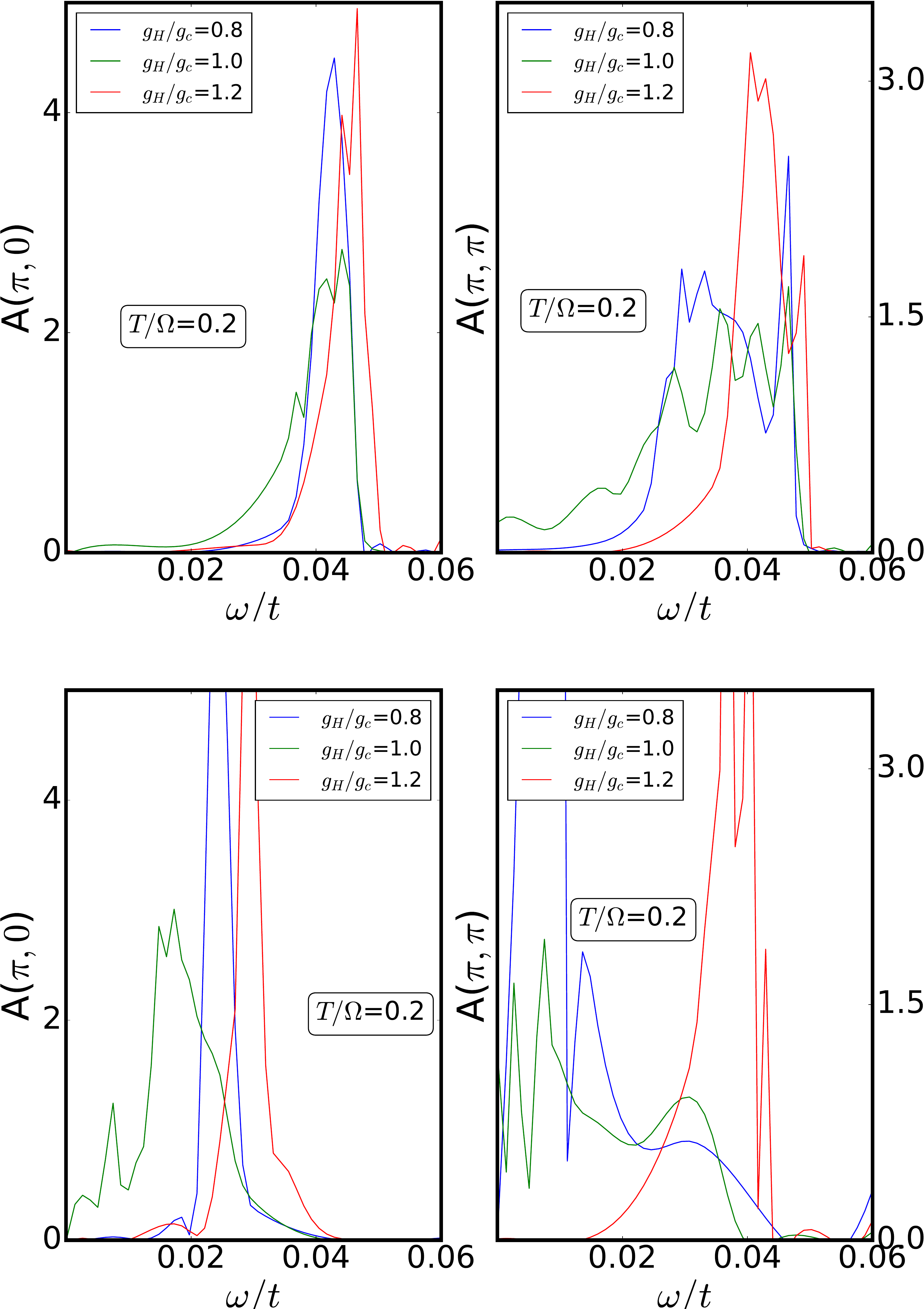}
}
\caption{Momentum resolved spectral functions
        $A(\pi,0)$ and $A(\pi,\pi)$ (left and right panels) of
        the optical (top row) and acoustic (bottom row) phonons
        for varying $g_{H}/g_c=(0.8,1.0,1.2)$. The relevant
        couplings along with color codes are
        listed. }
\end{figure}
% -----------------------------------------------------------
% -----------------------------------------------------------
\begin{figure}[t]
\centerline{
\includegraphics[height=3.5cm,width=8.5cm]{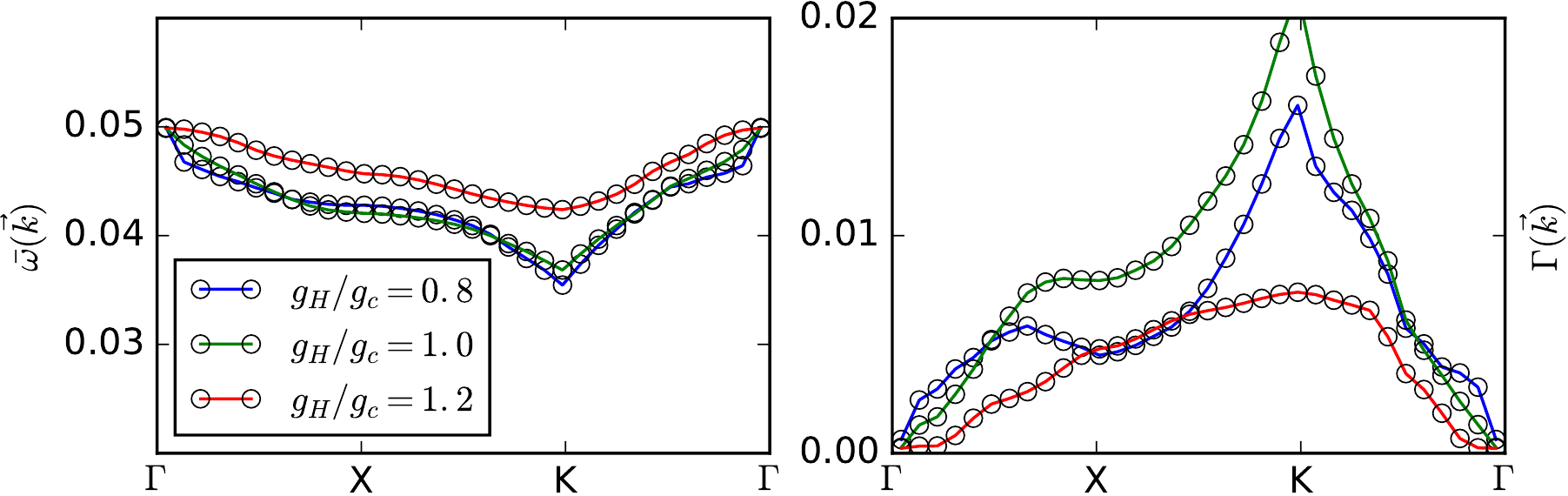}
}
\centerline{
\includegraphics[height=3.5cm,width=8.5cm]{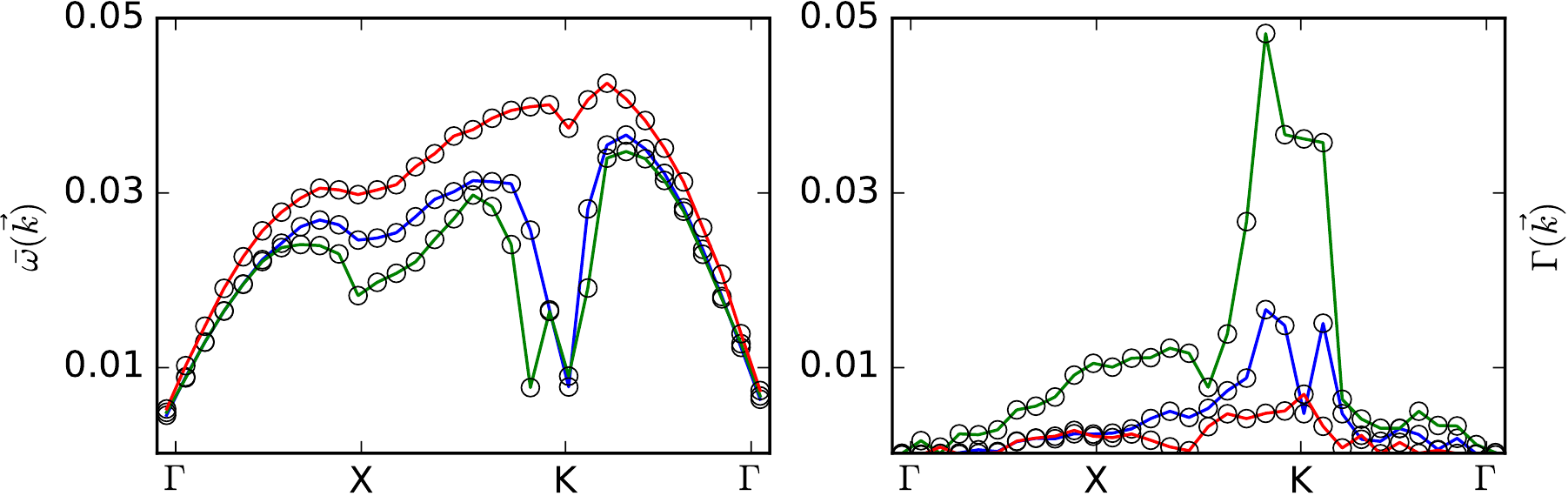}
}
\caption{Fitting of optical (top panel) and acoustic (bottom
        panel) phonon spectra for the ground state in terms of
        mean frequencies 
($\bar{\omega}(\vec{k})$ and linewidths ($\Gamma(\vec{k})$
        throughout the BZ trajectory
$(0,0)\rightarrow(\pi,0)\rightarrow(\pi,\pi)\rightarrow
        (0,0)$. The relevant couplings are listed along with
        color codes.}
\end{figure}
% -----------------------------------------------------------

\begin{equation}
\tilde{\Pi}_{ab}={A}_{ai} {\Pi}_{ij} {B}_{jb}
\end{equation} 
where $\Pi_{ij}$ is the same
 as in equation (14). The matrices 
sandwiched to the left and 
right are comprised of weighted
 form factors. They have
  the following expressions-
\begin{eqnarray}
A_{ai}&=&\sum_{\vec{q}}\frac{|\vec{q}|}{\sqrt{2\mu
\omega_{\vec{q}}}}e^{i(\vec{r}_i-\vec{r}_a).\vec{q}} \nonumber \\
B_{jb}&=&\sum_{\vec{{q}^{\prime}}}
\frac{|\vec{{q}^{\prime}|}}{\sqrt{2\mu
\omega_{\vec{{q}^{\prime}}}}}e^{i(\vec{r}_j-\vec{r}_b).\vec{{q}^{\prime}}}
\end{eqnarray}
Finally, a Dyson equation 
is solved (in real space) 
as in the optical phonon case- 
\begin{equation}
[\tilde{D}]^{-1}_{ab}(\omega)=
[\tilde{D}]^{-1}_{0,ab}(\omega)-\tilde{\Pi}_{ab}(\omega)
\end{equation}

\section{Results on the ground state}

We organize the results as follows. First, the optimal static 
background features are shown. These include
individual MC snapshots, the distribution of static 
optical phonon distortions $P(x_{0})$
the static phonon structure factor along the BZ diagonal 
($S(q,q)$) and electronic density of states (DoS) in the
ground state. 
Next, the spectral maps in 
($\vec{k},\omega$) space of the optical and acoustic
phonon propagators, 
calculated with the full frequency dependent 
$\Pi_{ij}(\omega)$, are featured. These spectra 
were calculated for moderately large sizes (24$\times$24). 
To enable a closer look into phonon properties, we plot the 
lineshapes for a couple of high symmetry $k$-points in the BZ -
($\pi,0$) and ($\pi,\pi$). 

% -----------------------------------------------------------
\begin{figure}[b]
\centerline{
\includegraphics[height=3.5cm,width=4.0cm]{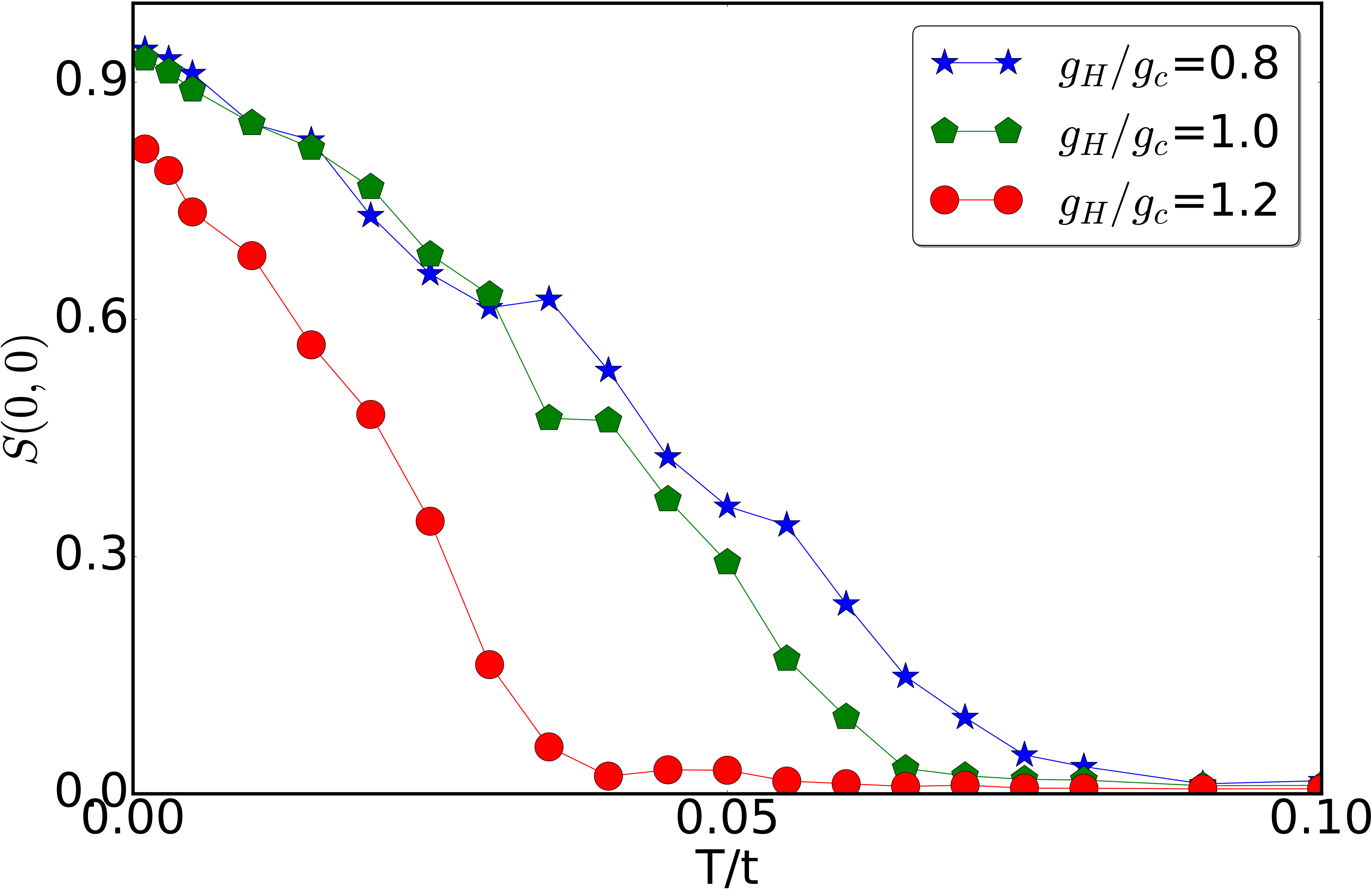}
\includegraphics[height=3.5cm,width=4.0cm]{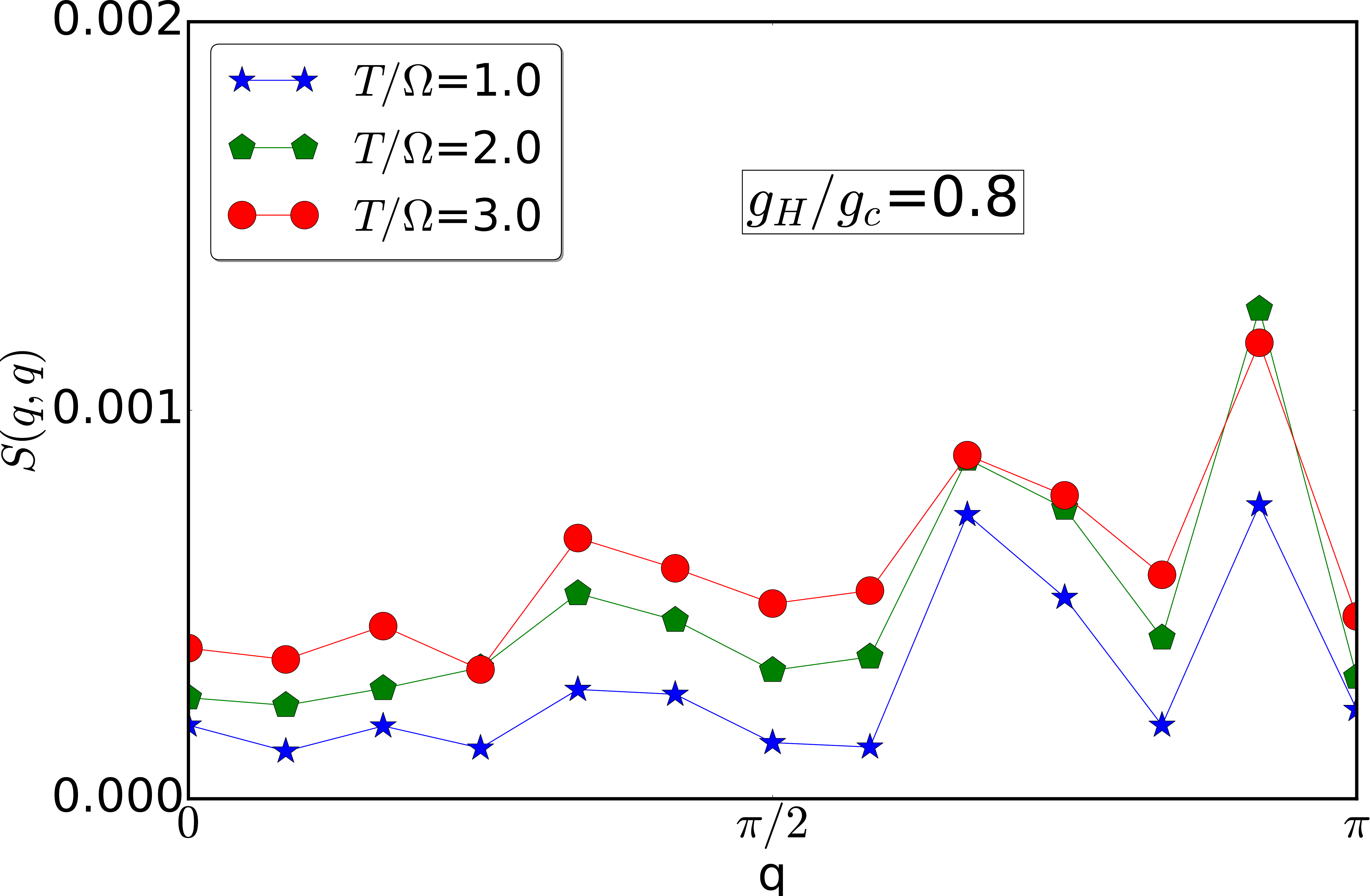}
}
\centerline{
\includegraphics[height=3.5cm,width=4.0cm]{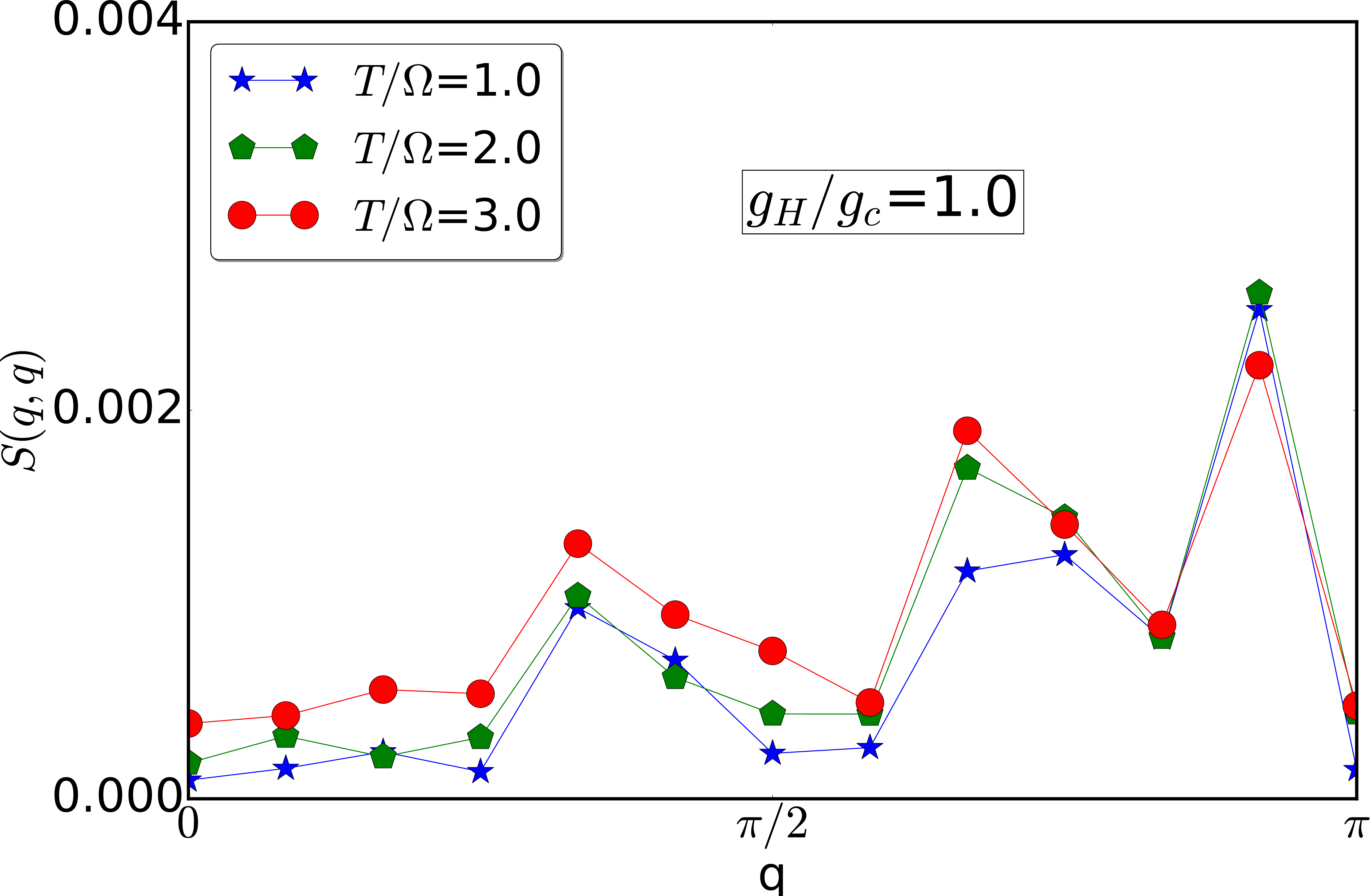}
\includegraphics[height=3.5cm,width=4.0cm]{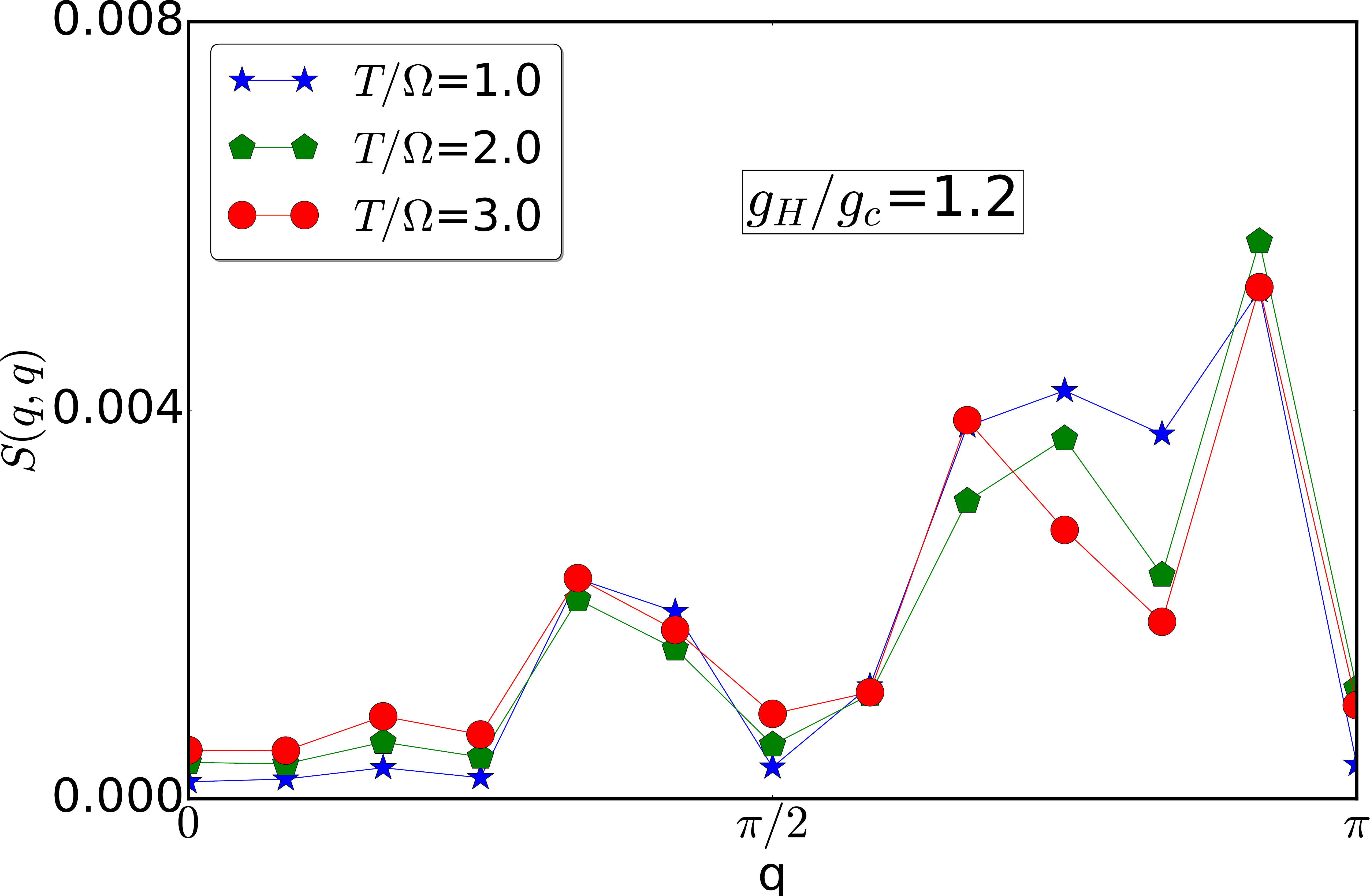}
}
\caption{Ferromagnetic structure factor $S(0,0)$
	(top left) for three couplings of interest
	and static phonon structure factors $S(q,q)$
	along the diagonal of BZ for three temperatures
	$T/\Omega=(1.0,2.0,3.0)$ in subsequent panels.}
\end{figure}
% -----------------------------------------------------------
% -----------------------------------------------------------
\begin{figure*}[t]
\centerline{
\includegraphics[height=4cm,width=14.5cm]{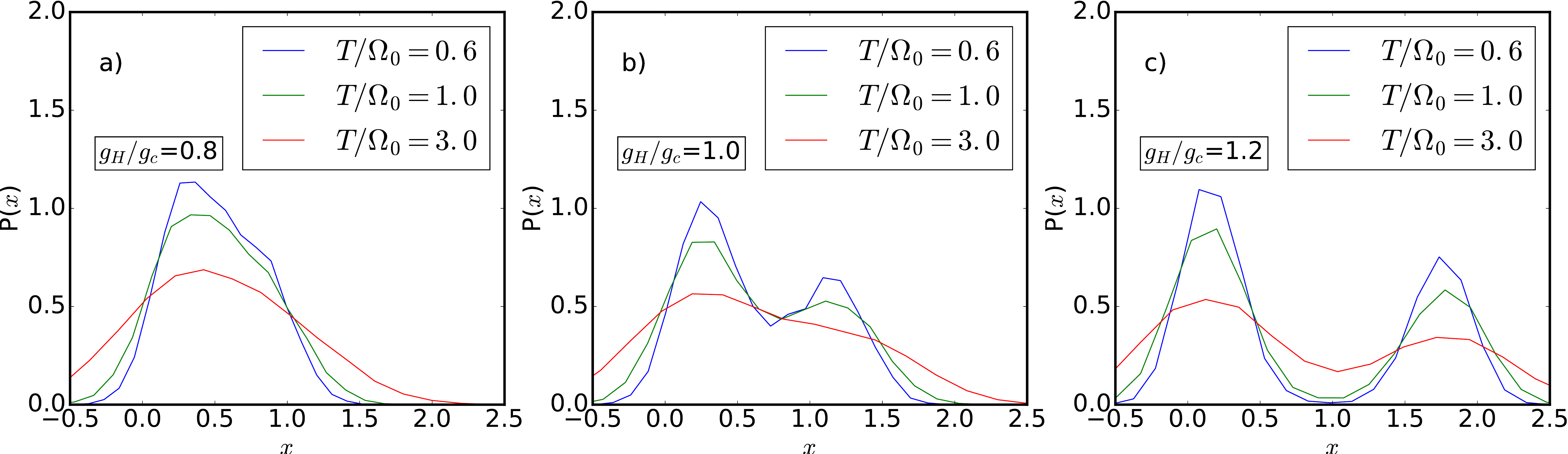}
}
\centerline{
\includegraphics[height=4cm,width=14.5cm]{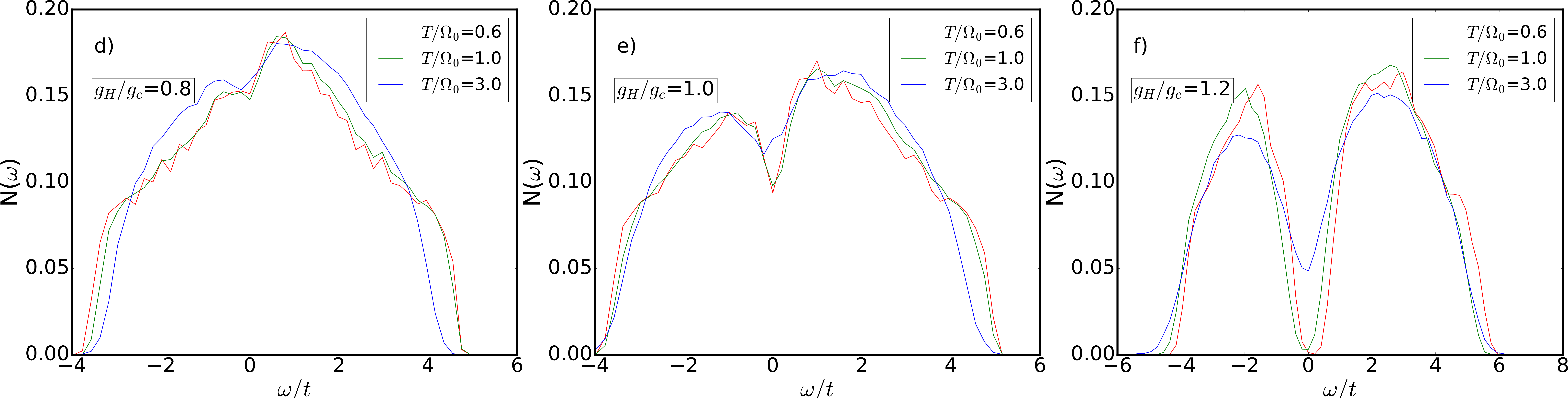}
}
\caption{Top row: The distribution of static optical mode
 distortions ($P(x)$) across three coupling windows
 a) weak ($g_{H}/g_c=0.8$), b) intermediate ($g_{H}/g_c=1.0$)
        and c) strong ($g_{H}/g_c=1.2$). The weak
        bimodality in the ground state at weak coupling
        smoothens out due to thermal
        fluctuations. At intermediate coupling, bimodality remains
        upto mild heating ($T/\Omega_0 \sim 1.0$). The strong
        coupling distribution is bimodal even upto
        $T/\Omega_0=3.0$. Bottom row: electronic DoS across the
        same couplings- d) a mild pseudogap seen at weak coupling
        for high $T$, e) intermediate coupling
        features gradually diminishing pseudogaps on heating,
        f) strong coupling DoS is gapped at
low $T$, has more robust pseudogap feature on heating.}
\end{figure*}
% -----------------------------------------------------------

Given that it's difficult to interpret 
these lineshapes directly, we've extracted mean
 frequencies and widths for the entire optical and acoustic 
 phonon spectra. We also attempt to make a connection between 
 the observed phonon features and the 
corresponding electronic physics on the static backgrounds. 
 The quantity which affects phonon features 
 directly in our theory is the polarization 
   $\Pi_{ij}(\omega)$. We don't provide any analysis
   of these in the present paper, but interpreting
   them is currently in progress.

\subsection{Backgrounds}

In this part, we show quantities characterizing the 
static optical mode which controls most of the 
phonon physics. First, the actual MC snapshots 
of lattice distortions are featured, followed by probability
distribution of these ($P(x)$), which indicate polaron
formation. Next, the
structure factor along the BZ diagonal shows emergence
of short-range correlations near ($\pi,\pi$). Finally, 
the electronic DoS in
 the ground state are exhibited, that
  show a gap formation at strong coupling. 

The MC snapshots, shown in top row of Fig.2, reveal 
static phonon correlations across various 
coupling windows. These pictures are representative 
of the ground state ($T/\Omega_0=0.2$). One sees a
 gradual improvement of $(\pi,\pi)$ correlations 
 amongst the distortions as we move from weak
  to strong coupling (left to right). However, at 
  weaker $g_{H}/g_c$, the features are much less underlined. 
  In the clean Holstein model, the strong coupling 
  state is supposed to be ordered, at least close to
   half-filling. Here, the weak binary disorder has
    the effect of destroying that, but short range 
    correlations persist. The temperature scale for 
    killing off these effects is set by $T_{CO}\sim\frac{t^2}{E_p}$.
     
% -----------------------------------------------------------
\begin{figure*}[t]
\centerline{
\includegraphics[height=11.0cm,width=9.50cm]{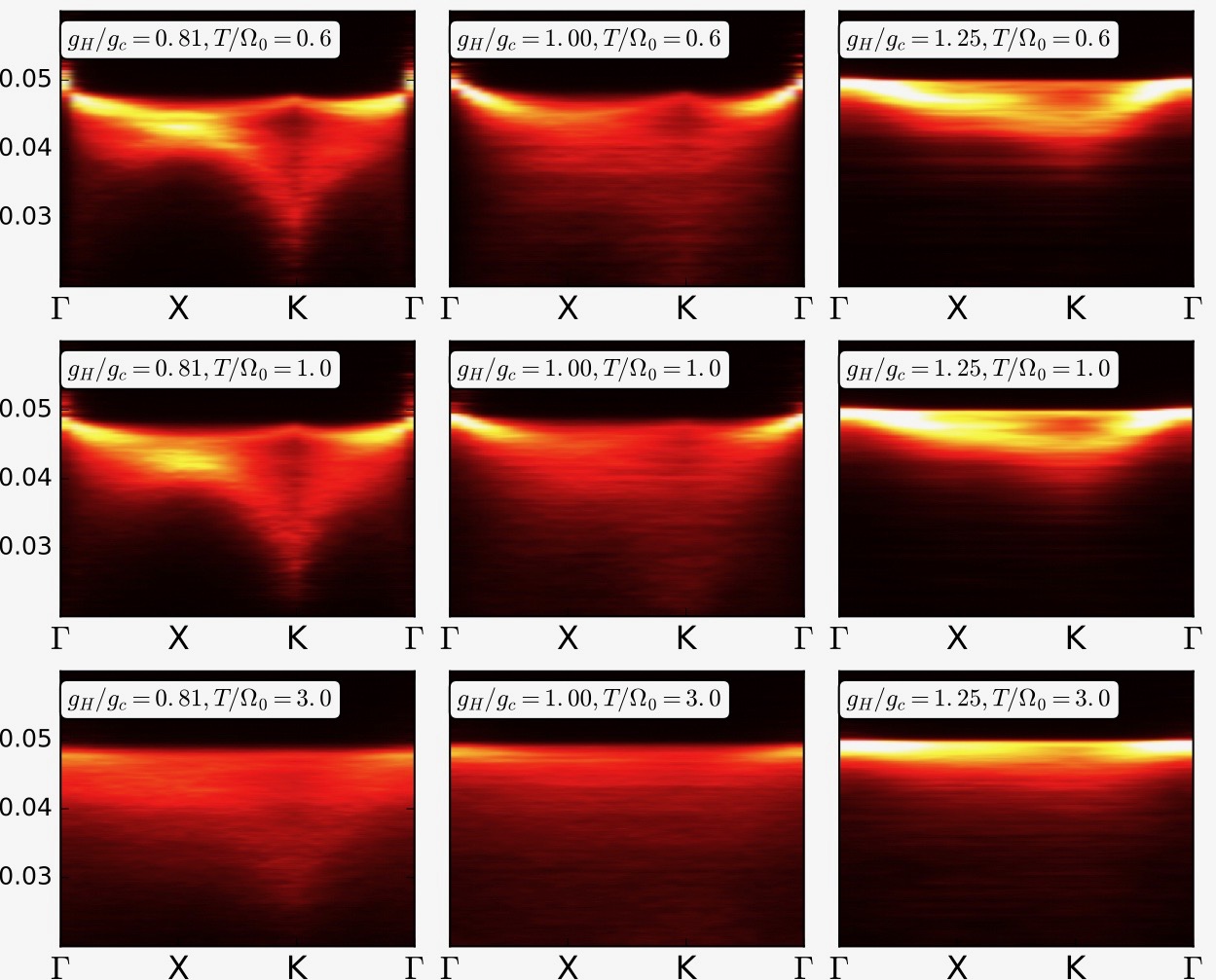} 
\includegraphics[height=11.0cm,width=10.5cm]{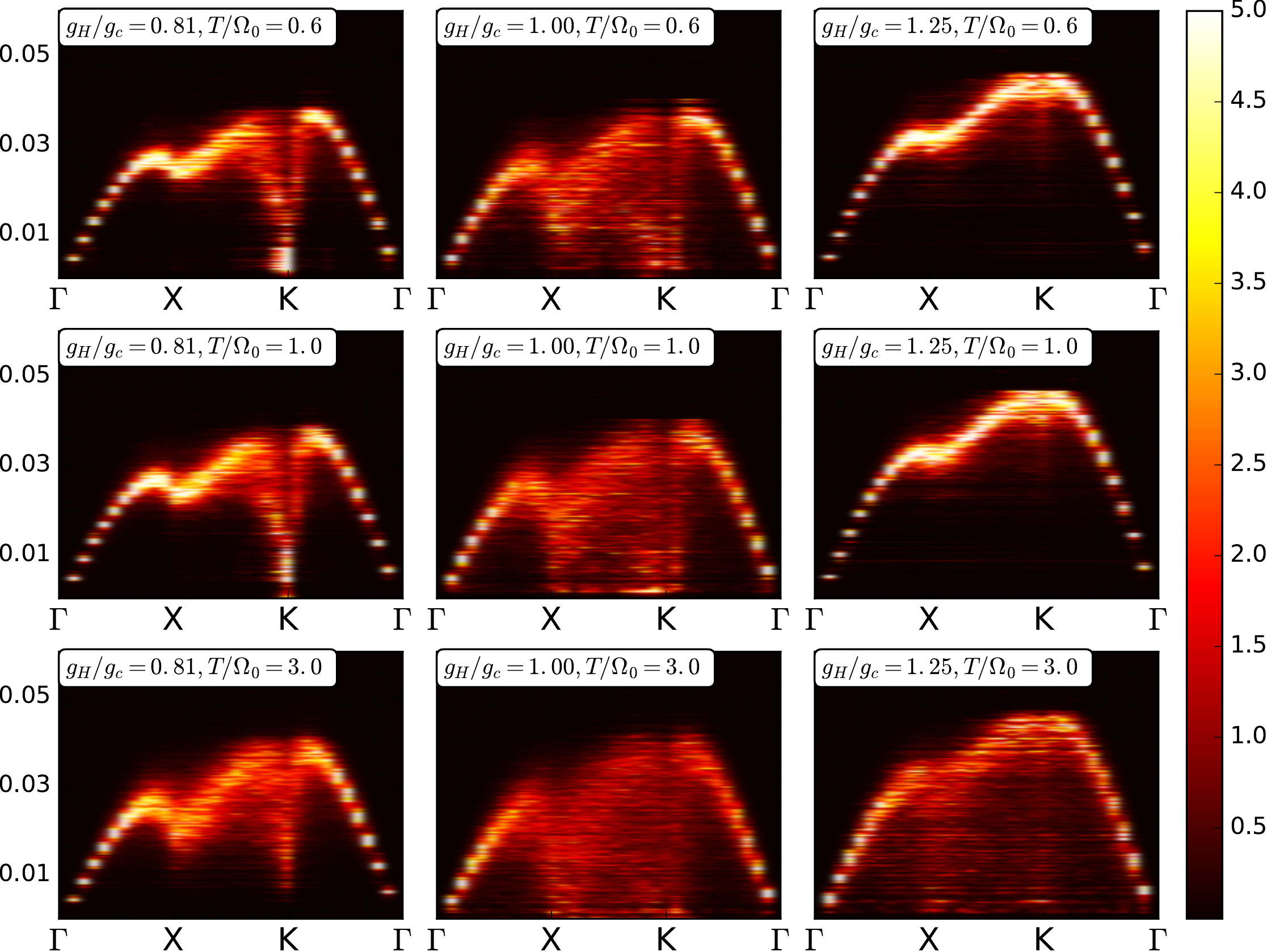} 
}
\caption{Left set: Spectral distribution of the optical
phonon Green's function for varying
$g_{H}$ and temperature. The trajectory
chosen along BZ is $(0,0)
\rightarrow(\pi,0)\rightarrow(\pi,\pi)\rightarrow
(0,0)$. Coupling increases along the
horizontal axis in the following sequence
(from left to right)- $g_{H}/g_{c}=(0.81, 1.00, 1.25)$.
Temperature varies form top to bottom-
$T/\Omega_0=(0.6,1.0,3.0)$.
The critical coupling at this density is $g_{c}=1.60$.
Right set:
Spectral distribution of the acoustic
                phonon Green's function for varying
                $g_{H}$ and temperature. The trajectory
                chosen along BZ is $(0,0)
                \rightarrow(\pi,0)\rightarrow(\pi,\pi)\rightarrow
                (0,0)$. Coupling increases along the
                horizontal axis in the following sequence
                (from left to right)- $g_{H}/g_{c}=(0.81, 1.00, 1.25)$.
                Temperature varies form top to bottom-
                $T/\Omega_0=(0.6,1.0,3.0)$.
                The critical coupling at this density is $g_{c}=1.60$.
                For these plots, $g_{ac}=g_{H}$.
}
\end{figure*}
% -----------------------------------------------------------

     The bottom row of Fig.2 shows first the distribution 
     of static distortions. At the weakest coupling, 
     a hint of bimodality is seen amidst a generally 
     broad picture.
     On going closer to the actual crossover ($g_{H}/g_c=1.0$), 
     a genuine two-peak distibution emerges with more 
     weight on the smaller $x$ value. At strong coupling, 
     a clear gap separates the two modes. The effect of 
     disorder is to enhance polaronic features at
     weaker couplings and broaden the peaks
     at strong coupling. 
     
     Next, we come to
     static phonon structure factor $S(q,q)$ in
      the ground state, from which the conclusions drawn
       from individual configurations can be further 
       corroborated. Close to the zone boundary, we see mild
       peaks, whose weights become prominent at strong
       coupling. 
       
      Finally, the electronic DoS also responds to the polaronic
      transition directly. We have a tight-binding like
      spectrum at weak coupling. The phonon induced
      renormalizations are small. Close to the transition,
      the behaviour changes qualitatively and a pseudogap
      emerges. At even higher couplings, polarons tend to order.
      As a result, we see formation of a small, hard gap.

\subsection{Overview of phonon modes}

In Fig.3, we exhibit the spectral maps of the optical 
and acoustic phonons in various coupling regimes. 
We already depicted such maps in our previous 
work$\cite{sauri}$ on the clean Holstein model. The
 first row is depicting optical phonons on the ground 
 state. Below $g_{H}=g_c$, the mode 
 is fairly sharp except in the vicinity of $(\pi,0)$, 
 and most noticeably 
 around $(\pi,\pi)$, where it also splits and softens to 
 some degree. We ascribe this behaviour to the proximity 
 to a nesting instability ($n=0.4$ is close to half-filling). 
 Compared to the clean Holstein problem, 
the broadenings are considerably large in the present case. The reason is the
 presence of weak binary disorder, which also affects the 
 Kohn anomaly like softening feature (near ($\pi,\pi$)) markedly. 

% -----------------------------------------------------------
\begin{figure}[b]
\centerline{
\includegraphics[height=7cm,width=8.5cm]{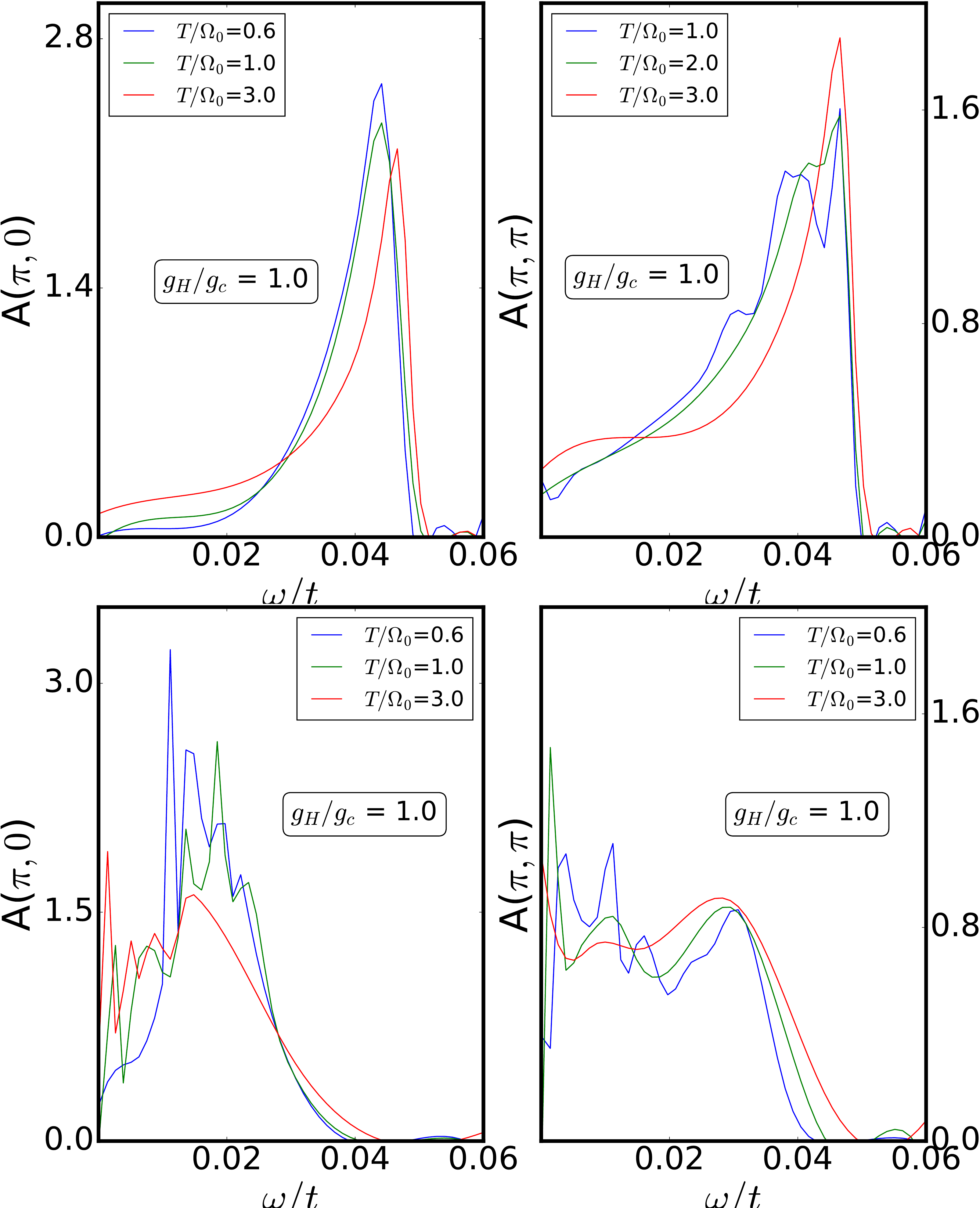}
}
\caption{Lineshapes of optical (top panel) and acoustic
        (bottom panel) phonons at $(\pi,0)$ and $(\pi,\pi)$
        for intermediate coupling ($g_{H}/g_c=1.0$).
        The relevant temperatures along with color codes
        are listed.}
\end{figure}
% -----------------------------------------------------------
% -----------------------------------------------------------
\begin{figure}[t]
        \centerline{ 
                \includegraphics[height=9cm,width=8.5cm]{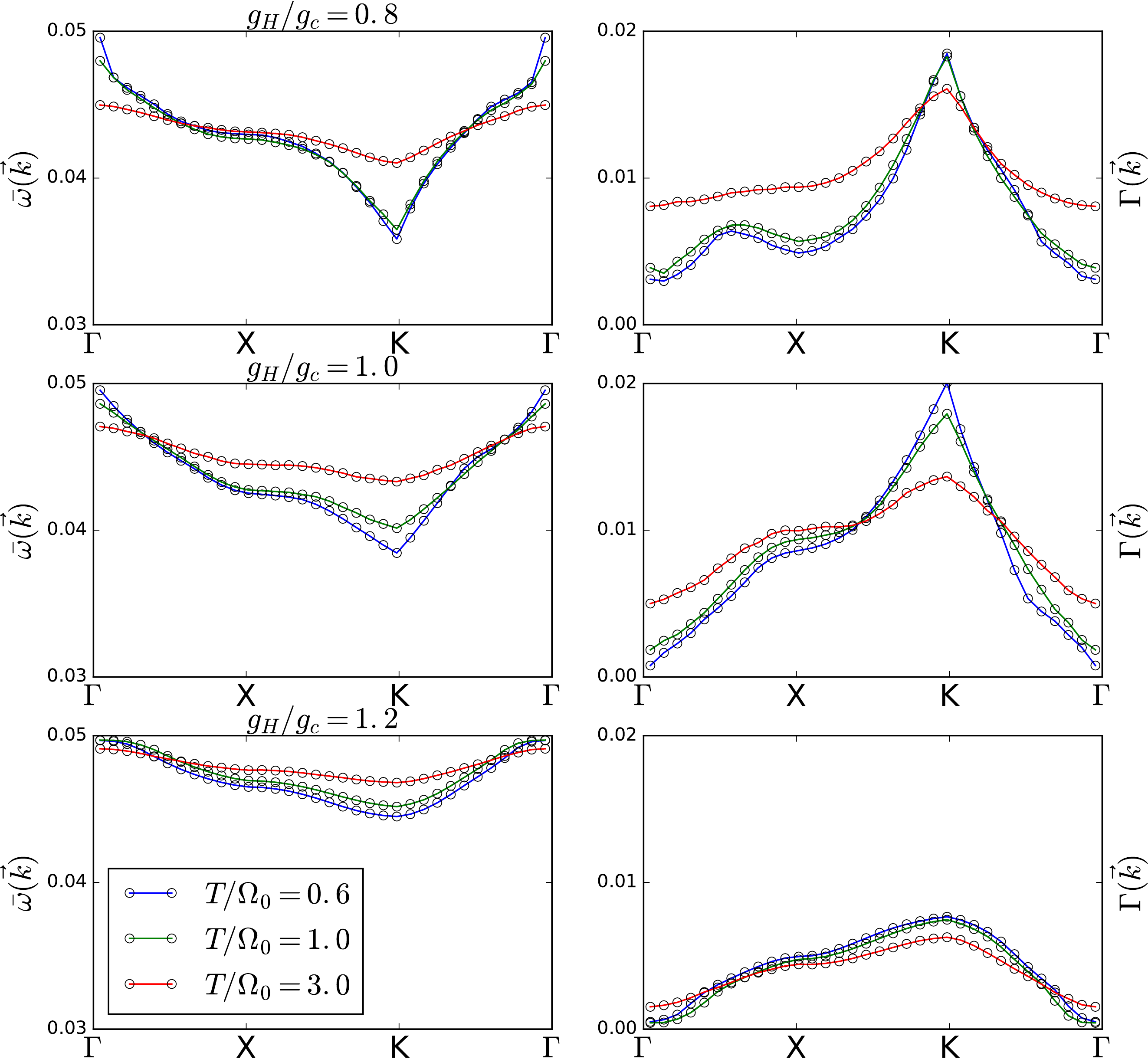}
        }
        \caption{Fitting of optical phonon spectra at finite
                temperature in terms of
                mean frequencies ($\bar{\omega}(\vec{k})$) and
                linewidths ($\Gamma(\vec{k})$)
                throughout the BZ trajectory
$(0,0)\rightarrow(\pi,0)\rightarrow(\pi,\pi)
                \rightarrow(0,0)$. Each panel contains three
                temperatures- $T/\Omega_0=(0.6,1.0,3.0)$.
                Coupling increases
                from top to bottom in the following sequence-
                $g_{H}/g_c=(0.8,1.0,1.2)$.}
\end{figure}
% -----------------------------------------------------------

 At $g_{H}=g_c$, polarons appear and 
 the optical phonon spectrum becomes significantly
  broad due to presence of quasi-localized phonons. 
  Except modes near the zone center ($0,0$), where
   scattering is prohibited owing to phase space 
   constraints, a dramatic damping effect is seen. The
   softening is much less compared to the clean problem.
    
   Beyond the polaronic transition ($g_{H}/g_c=1.25$), 
  the phonon branch sharpens again and the softening
   is considerably less. We see a remnant of the branching 
  feature, observed prominently for the clean problem
  over a region in momentum space (from $X$ to $K$), owing
          to charge correlations and the resulting imperfect order.

 Moving to the second row, we discuss the acoustic phonon
  features now. Below $g_{H}=g_c$, the mode is clearly 
  dispersive except around $(\pi,\pi)$, where it softens 
  remarkably. We relate this feature, as in the 
case of optical phonons, to the proximity to a nesting instability 
  ($n=0.4$ in our study). We note that acoustic phonons
  respond to nesting physics much more dramatically, despite
  disorder effects being present. 
   
  At $g_{H}=g_c$, polarons formation causes the phonon 
  distribution to become significantly broad 
due to formation of large static distortions. 
Except modes near the zone center ($0,0$), where 
scattering is prohibited owing to the form of 
acoustic phonon-conduction electron coupling, a dramatic 
  softening and damping is seen. 
  
  Beyond the polaronic 
  transition ($g_{H}/g_c=1.25$), the phonon branch sharpens 
  again and the softening is considerably reduced. 
  States near the Fermi level are depleted in this regime 
  and hence the scattering is less.

\subsection{Detailed behaviour of phonon modes}

\subsubsection{Optical phonons}

The optical phonon lineshapes in the ground state (top
row of Fig.4) suggest that at $(\pi,0)$, we have a well-defined
mode at weak coupling ($g_{H}/g_c=0.8$). The spectrum evolves
to one having a long tail
towards lower frequencies close to the crossover. But,
on going further in the polaronic phase, the mode sharpens
once again, with a mild, split-peak feature.

The right panel in the top row exhibits the $(\pi,\pi)$ 
lineshape. This is broad even at weak coupling, with more
weight in lower frequencies. We ascribe this behaviour to 
presence of disorder and proximity to perfect nesting, as
discussed earlier. On moving close to the crossover 
($g_{H}/g_c=1.0$), the decoherence is more severe. Large static
distortions are deemed responsible for this. The spectrum
sharpens and hardens at strong coupling ($g_{H}/g_c=1.2$).
There's a hint of spectral splitting as well.

% -----------------------------------------------------------
\begin{figure}[b]
        \centerline{
                \includegraphics[height=9cm,width=8.5cm]{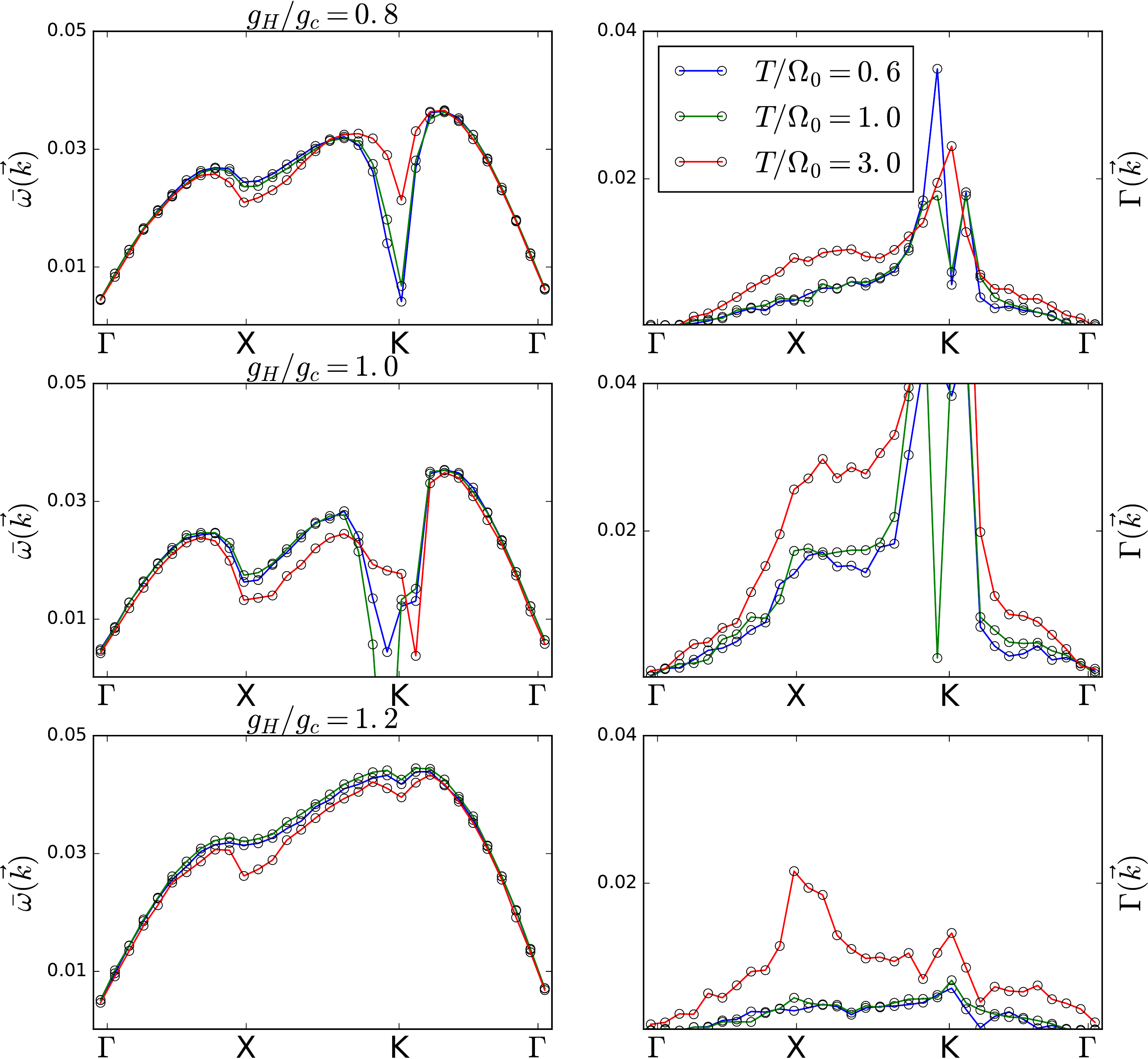}
        }
        \caption{Fitting of acoustic phonon spectra at finite
                temperature in terms of
                mean frequencies ($\bar{\omega}(\vec{k})$) and
                linewidths ($\Gamma(\vec{k})$)
                throughout the BZ trajectory $(0,0)\rightarrow(\pi,0)\rightarrow(\pi,\pi)
                \rightarrow(0,0)$. Each panel contains three
                temperatures- $T/\Omega_0=(0.6,1.0,3.0)$.
                Coupling increases
                from top to bottom in the following sequence-
                $g_{H}/g_c=(0.8,1.0,1.2)$.}
\end{figure}
% -----------------------------------------------------------

To gain further insight into the phonon features across
the full Brillouin zone, we've extracted mean frequencies
and linewidths along the BZ trajectory mentioned in the 
maps. These are provided in the top row of Fig.5. 
The softening (left panel) is clearly dominant at $(\pi,\pi)$, 
and decreases at strong coupling. The linewdiths also
feature a peak at the same momentum point. The signature is
most prominent at intermediate coupling, when polarons start
forming but short-range correlations are still nascent. 

\subsubsection{Acoustic phonons}

The acoustic phonon lineshapes on the ground state (bottom
row of Fig.4) show qualitatively similar trends
to the corresponding optical phonons at $(\pi,0)$ 
(left panel). One notable
difference lies in the fact that the mode softens instead
of hardening at intermediate coupling ($g_{H}/g_c=1.0$). 
In the deep polaronic phase, the spectrum shifts again
to higher frequencies and lifetime is bandwidth limited.

The $(\pi,\pi)$ lineshapes (right panel) are more non-trivial
to interpret. The weak coupling spectrum has a mode at
very low ($\omega/t=0.01$) frequency. This is connected
to Kohn anomaly. Additionally, a broad shoulder is observed,
owing to disorder effects. The coherence at low frequency
is lost out closer to the transition, where the static
optical mode has large distortions. Going to higher coupling
($g_{H}/g_c=1.2$), the mode is still broad but regains
sharp features near the bare frequency. A splitting 
character is observed similar to optical phonons.

In the bottom row of Fig.5, we plot the mean frequencies
and widths, as in case of optical modes. The undamped
Goldstone mode at $(0,0)$ is omitted for clarity. The sharp
dip of mean values near $(\pi,\pi)$ is underlined (left
panel). Corresponding linewidths, interpreted as lifetimes,
increase markedly near the transition ($g_{H}/g_c=1.0$). 
We relate this to emerging short-range correlations amongst
polarons. Once the ordering sets in at strong coupling
($g/g_c=1.2$), widths decrease again. 

\section{Effect of thermal fluctuations}

After discussing phonon properties for the ground
state and relating them
to background and electronic features, we now move
to the thermal physics. Fig.1 depicts the thermal
phase diagram in our model for high ($n\sim0.4$)
denisties. The FM and PM are ferromagnetic and
paramagnetic metallic phases. The corresponding 
insulating phases are FI and PI, respectively. 
The blue-lettered regions are phonon regimes.
In the weak coupling regime, standard
perturbation theory results for phonons are
useful. In the fan-like crossover region, phonons 
are very broad ($\Gamma\sim\Omega_{0}$). Towards the
right, we have narrow band optical phonons. Compared
to the clean Holstein phase diagram\cite{sauri}, 
we see the region featuring electronic pseudogap 
and `crossover' phonons have shifted to the left.
The electronic phases are indicated in black letters.

\subsection{Backgrounds}

The ground state of our model is ferromagnetically ordered
at all coupling windows, owing to double exchange interaction.
We first plot the $S(0,0)$ structure factor to infer
$T_c$ scales (top left panel of Fig. 6). 
At weak coupling $g_{H}/g_c=0.8$,
the $T_c\sim0.07t$. This value reduces to $T_c\sim0.04t$
at strong coupling ($g/g_c=1.2$). The ferromangetic $T_c$
plays an important role in the broadening of phonons
as the interplay of spin disorder and thermal fluctuations
has direct impact on the electronic spectrum. This in turn
feeds back to the phonons through the polarization 
$\Pi_{ij}(\omega)$ within our calculation.

The top right and bottom (left and right) panels of 
Fig.6 show static phonon structure factors 
(shown for the ground state in
Fig.2, middle panel) at finite temperature. The rising 
feature near ($\pi,\pi$) survives and in fact grows
mildly on heating at weak coupling.

In Fig.7, the first row depicts static distributions of 
optical mode distortions in various coupling regimes. The
left panel features weak coupling regime, where the
hint of bimodality observed in the ground state is wiped
out on mild heating. Increasing the temperature further 
(going beyond ferro $T_c\sim0.07t$), we see a very broad
distribution skewed to lower $x$ values. The intermediate
coupling (middle panel) picture retains bimodality upto 
the bare phonon scale $T/\Omega_0=1.0$. At strong coupling,
even at high $T$, we have a two-peak feature.

In the bottom row (Fig.7), thermal dependence of electronic
DoS is shown. The most non-trivial feature is a weak hint of
pseudogap at weak coupling (left panel). This arises out
of spin disorder, causing effective electron hopping to
weaken. There are also large thermal fluctuations of
the static mode, which grow into actual distortions
at intermediate coupling ($g_{H}/g_c=1.0$). There's a
pseudogap surviving upto high temperature in the middle
panel (close to crossover). At higher coupling (right panel),
the lowest $T$ features a small gap, that develops into
a robust pseudogap on heating up.  

\subsection{Overview of phonon modes}

\subsubsection*{Optical phonons}

In Fig.8 (left set), we observe the spectral maps of
 the optical (left) and acoustic (right) phonons 
 in various coupling
  and temperature regimes. The first row 
  is representative of phonons at low $T$ 
  ($T/\Omega_0=0.6$). Ferromagnetic
  order is still present in the background.
  Below $g_{H}=g_c$, the
   mode is still `dispersive' except around $(\pi,\pi)$, 
   where the softening feature is retained. At $g_{H}=g_c$, 
   we now have thermal fluctuations of 
       large static distortions playing out
       in tandem with magnetic fluctuations. 
       Except modes near the 
       zone center ($0,0$), a 
       dramatic softening and damping is seen.
       The high temperature phonons are very broad ($\Gamma\sim 
\Omega$). 
       Beyond the polaronic transition ($g_{H}/g_c=1.25$),
        the phonon branch sharpens again and the softening
         is considerably reduced. The mild branching
         feature observed near ($\pi,\pi$) for the ground state
         is washed out. The role of disorder
         is to effectively promote electronic spectral
         gaps. Consequently, these phonons are less
         broad compared to those in the clean Holstein model.

\subsubsection*{Acoustic phonons}

Moving to the spectral maps of the finite temperature
acoustic phonons (Fig.8, right set), at the scale 
of bare Holstein frequency ($T/\Omega_0=1.0$), we observe 
quantitative changes at weak and 
strong coupling but a significant increase in lifetimes 
for the middle (intermediate coupling) picture. At even 
higher temperatures, well past the ferro $T_c$ 
scale for all couplings, phonons in the polaronic side are rendered 
quite broad and incoherent, except at low wavevectors. 
The metallic side ($g_{H}/g_c\sim0.8$) is affected mostly 
near $(\pi,\pi)$, where the degree of softening reduces
and lifetime picks up. 

\subsection{Detailed behaviour of phonon modes}

As in case of the ground state, we study the phonons more
closely by examining specific lineshapes and fitting the
full spectrum to find mean frequencies and linewidths.
Here, we focus on the most interesting regime of
intermediate coupling ($g_{H}/g_c=1.0$).

\subsubsection*{Optical phonons}

In this regime, the polaronic crossover 
causes both the ($\pi,0$) and ($\pi,\pi$) lineshapes 
to have long tails at all temperature windows, as seen in 
Fig.9 (top panel). The effect of large static distortions 
is most prominent in the ($\pi,\pi$) lineshapes. Some features 
of sharp peaks remain at low $T$, owing to limited 
system size ($24\times24$) which gradually smudge out, 
giving an incoherent spectrum with some weight near 
the unrenormalized frequency on further heating 
($T/\Omega_0=3.0$). One may ascribe this behaviour partly 
to loss of ferromagnetic order also, which takes place 
around $T=0.06t$.

Fig.10 shows the mean frequencies (left column) and
linewidths (right column) as functions of temperature
for three regimes. The
trends mentioned previously are clealy visible. First, 
at `weak' coupling, proximity to a nesting instability 
creates phonon softening around the zone boundary. 
broadenings also pick up in that region. On heating 
up across the ferromagnetic $T_c\sim0.07t$, a marked 
increase in lifetimes take place. The momentum 
dependence of the spectrum is also subdued. 
Moving to the polaronic crossover regime, the 
ground state picture is roughly unaltered, 
whereas thermal broadenings generally decrease, 
due to formation of short-range correlations. 
This effect is markedly observed moving to the 
strong coupling (third column) plots. The mean 
(or mode) frequencies harden and lifetimes become
limited by the bandwidth, which scales as $\frac{t^2}{E_p}$. 

\subsubsection*{Acoustic phonons}

At intermediate coupling, the polaronic crossover imparts 
an overall damping of acoustic phonon spectrum throughout 
the Brillouin Zone (BZ). However, the effect is most drastic 
around $(\pi,\pi)$, where across a wide range of temperatures,
a `box-like' lineshape emerges (Fig.10, bottom panel). 
The thermal dependence of lineshapes is not very significant. 
Even at $(\pi,0)$, there are large lifetimes present 
along with prominent softening, especially at high T 
($T/\Omega_0=3.0$).

The corresponding fitting plots of the acoustic phonons are 
exhibited in Fig.11. The left column, depicting the 
mean frequencies, show prominent dips at both $(0,0)$ 
and $(\pi,\pi)$ in the weak coupling regime. 
The Goldstone mode at $\Gamma$ point is undamped (not shown 
explicitly). The softening of phonon mode at $(\pi,\pi)$
owes its origin to Kohn anomaly. Close to the crossover, 
broadenings increase quantitatively, due to formation of 
large static distortions and consequent phonon localization.
Thermal fluctuations magnify the effect remarakably past
the ferromagnetic $T_c\sim0.06t$. Low momentum modes 
are protected from scattering because of phase space 
restrictions. At strong coupling, the dispersion
regains clarity to a large degree. Only on heating up to 
$T/\Omega_0=3.0$, one observes significant broadenings, 
near $(\pi,0)$. 

% -----------------------------------------------------------
\begin{figure}[b]
        \centerline{
                \includegraphics[height=3.5cm,width=8.5cm]{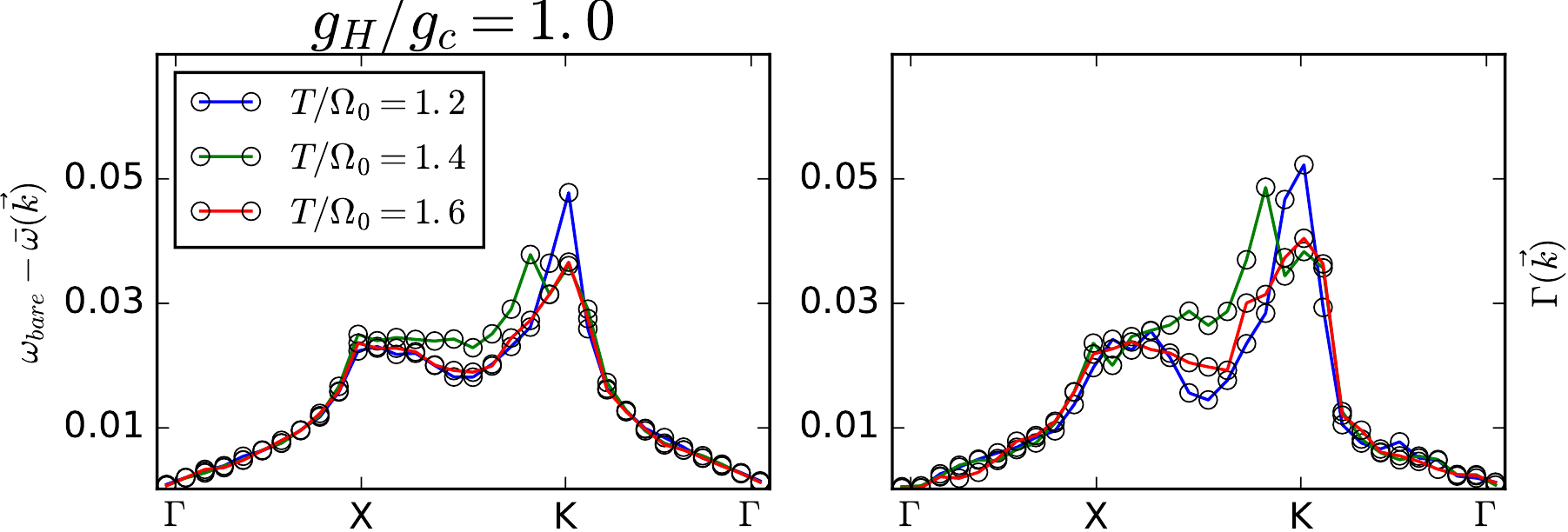}
        }
        \caption{Fitting of acoustic phonon spectra at finite
                temperature in terms of `softenings'
                (deviation of mean frequencies ($\bar{\omega}(\vec{k})$
                form bare values) and linewidths ($\Gamma(\vec{k})$
                through the BZ trajectory $(0,0)\rightarrow(\pi,\pi)$.
                The coupling is $g_{H}/g_c=1.0$.
                Temperature increases
                in the following sequence-
                $T/\Omega_0=(1.2,1.4,1.6)$. The
                ferromagnetic transition occurs at $T_c=1.2\Omega_{0}$}.
\end{figure}
% -----------------------------------------------------------
% -----------------------------------------------------------
\begin{figure}[t]
	\centerline{
		\includegraphics[height=9cm,width=8.5cm]{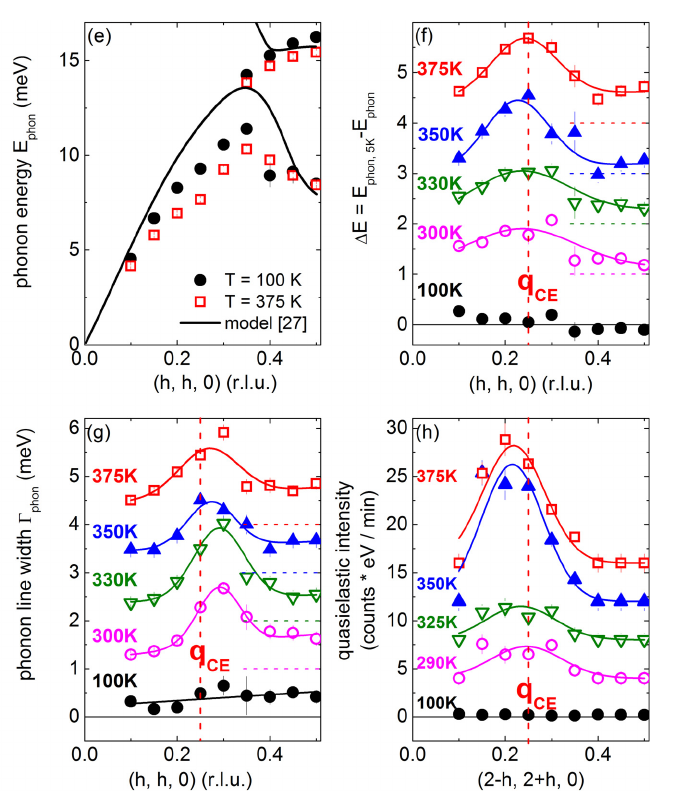}
	}
	\caption{Summary of investigations on transverse acoustic
		phonons in La$_{0.7}$Sr$_{0.3}$MnO$_{3}$. In
		panel (e) (top left), we have phonon energies in
		meV at two different temperatures ($100K$ and $375 K$).
		Also, a model calculation is shown. In panel (f)
		(top right), phonon softenings compared to low$T$
		($5K$) values are featured for several temperatures
		along the $(h h 0)$ direction in BZ. The $T_c=350K$
		marks the onset of pronounced softenings around 
		$\vec{q}_{CE}$. The bottom left panel (g) shows
		corresponding linewidths, which pick up just before
		the metal-metal transition. Lastly, in panel (h) 
		(bottom right), one has quasi-elastic intensitites
		quantifying short-range ordering of polarons beyond 
		$T_c$.
	}
\end{figure}
% -----------------------------------------------------------

\section{Discussion}

We list below the salient features of
 the phonon modes revealed through our investigation-

\begin{itemize}
\item The low temperature state of 
our model is ferromagnetically ordered and 
shows charge correlations at strong coupling. 
\item The weak coupling `disordered 
metal' at low $T$ is in proximity to a nesting
 instability for $n\sim0.5$. As a result, both 
 optical and acoustic modes are softened 
 considerably around $(\pi,\pi)$. 
\item On heating up the weak coupling state, two 
effects occur simultaneously. Firstly, thermal 
fluctuations induce large static distortions in the
 optical mode. Secondly, the ferromagnetic order
  is lost as one goes across $T_c\sim0.1t$. 
  This leads to broad, flat optical phonon 
  spectrum and a loss of coherence (barring 
  small wavevectors) in the acoustic mode.
These features are correlated with the
 appearance of a weak pseudogap in the 
 electronic density of states, owing to 
 spin disorder.
\item The intermediate coupling state is on
 the verge of a polaronic transition. So,
  phonon eigenstates tend to get localized. 
  This results in generally broad phonon modes 
  except around $(0,0)$. On increasing 
  temperature, the lifetimes increase quantitatively, 
  along with a flattening of the optical phonons. 
  The electronic spectrum is already pseudogapped
   in the ground state, the extent of which
    becomes milder on heating up. 
\item At stronger couplings, a short-range
 $(\pi,\pi)$ order sets in amongst polarons 
 at low $T$. As a result, the electronic spectrum
  shows a small hard gap. There's a depletion 
  of states near Fermi level, leading to weaker 
  low-energy weights in the polarization
   based self-energy for the phonons. 
   Consequently, both acoustic and 
   optical modes regain some coherence. 
   The damping is bandwidth limited
    $\Gamma\sim\frac{t^2}{E_p}$.
\item Heating up and going across the
 ferro $T_c$ (which is almost half of
  its weak-coupling value), one creates
   a more disordered background for the 
   phonons to propagate. This causes
    broadening effects to pick up. The electronic 
    spectrum shows a prominent pseudogap 
    and the acoustic phonons are
     broadened considerably. 
\end{itemize}

We've tried out a qualitative comparison with 
actual experimental data for acoustic phonons in
La$_{0.7}$Sr$_{0.3}$MnO$_{3}$\cite{weber2}. Fig. 
12 summarizes our results. The general
trends show an agreement. Our acoustic phonons
show increased softening and broadening effects
at intermediate coupling ($g_{H}/g_c=1.0$)
just before the ferromagnetic transition $T_c\sim1.2\Omega$
at $K$ point. This is also the important wavevector
for short-range ordering of polarons. Correspondingly,
in the experimental plots (Fig. 13), we see around 
$\vec{q}_{CE}$, the softening and linewidths pick up, 
just before $T_c=350K$. The increase is monotonic in 
both cases along the BZ trajectories chosen. 
The quasielastic intensity, which
quantifies short-range ordering of distortions,
shows a marked increase near the same wavevector
at $T_c$. On crossing this temperature, both
the theoretical and experimental curves show
a mild reduction in broadenings.

\section{Conclusions}

In conclusion, we've studied the phonon spectrum of
the weakly disordered
double exchange Holstein model in two dimensions.
The results have direct implications for recent
inelastic neutron scattering experiments in manganites.
The novelty of the method is to be able to capture
short-range correlations amongst polarons and
resulting pseudogap features in the electronic
spectrum at finite temperature. This in turn affects
the polarization of the system, which renormalize
the acoustic and optical phonons. The most interesting
region in our parameter space is one of intermediate
coupling and temperature, when polarons start forming 
and the ferromagnetic order in the ground state is
on the verge of destruction. We've quantified the
observed spectrum through detailed fitting across BZ
and an analysis of lineshapes at specific high-symmetry
points. We are investigating the problem at even weaker
values of binary disorder, so as to disentangle the polaron
physics from the effect of disorder. We also wish
to compare our findings with experiments in more detail 
in the future.

{\it Acknowledgement:}
We acknowledge the use of HPC clusters at HRI. SB
acknowledges fruitful discussions with Samrat Kadge
and Abhishek Joshi, and thanks Samrat Kadge for a careful
reading of the manuscript. The research of SB was partly
supported by an Infosys scholarship for senior students.
The research of SP was partly supported by the  Olle
Engkvist Byggm\"{a}stare Foundation.

\bibliographystyle{unsrt}

\end{document}